\documentclass[twocolumn]{aastex63}

\usepackage{xcolor}

\def\JB{{\rm Jy~beam^{-1}}}
\def\mJB{{\rm mJy~beam^{-1}}}
\def\uJB{{\rm \mu Jy~beam^{-1}}}

\def\kms{{\rm km~s^{-1}}}
\def\Msun{M_{\sun}}

\revised{\today}
\submitjournal{ApJ}

\graphicspath{{./}{figures/}}

\shorttitle{Polarization and spiral in TMC-1A}
\shortauthors{Aso et al.}

\begin{document}

\title{Multi-scale Dust Polarization and Spiral-like Stokes-I Residual in the Class I Protostellar System TMC-1A}

\correspondingauthor{Yusuke Aso}
\email{yaso@kasi.re.kr}

\author[0000-0002-8238-7709]{Yusuke Aso}
\affil{Korea Astronomy and Space Science Institute (KASI), 776 Daedeokdae-ro, Yuseong-gu, Daejeon 34055, Republic of Korea}

\author[0000-0003-4022-4132]{Woojin Kwon}
\affil{Department of Earth Science Education, Seoul National University, 1 Gwanak-ro, Gwanak-gu, Seoul 08826, Republic of Korea}

\author[0000-0001-9304-7884]{Naomi Hirano}
\affil{Academia Sinica Institute of Astronomy and Astrophysics, 11F of ASMA Building, No.1, Sec. 4, Roosevelt Rd, Taipei 10617, Taiwan}

\author[0000-0001-8516-2532]{Tao-Chung Ching}
\affil{CAS Key Laboratory of FAST, National Astronomical Observatories, Chinese Academy of Sciences, People{\it '}s Republic of China}
\affil{National Astronomical Observatories, Chinese Academy of Sciences, A20 Datun Road, Chaoyang District, Beijing 100012, People{\it '}s Republic of China}

\author[0000-0001-5522-486X]{Shih-Ping Lai}
\affil{Department of Physics, National Tsing Hua University, 101 Section 2 Kuang Fu Road, 30013 Hsinchu, Taiwan}
\affil{Institute for Astronomy, National Tsing Hua University,101 Section 2 Kuang Fu Road, 30013 Hsinchu, Taiwan}
\affil{Academia Sinica Institute of Astronomy and Astrophysics, P.O. Box 23-141, 10617 Taipei, Taiwan}

\author[0000-0002-7402-6487]{Zhi-Yun Li}
\affil{Department of Astronomy, University of Virginia, Charlottesville, VA 22904, USA}

\author[0000-0002-1407-7944]{Ramprasad Rao}
\affil{Center for Astrophysics ${\it |}$ Harvard \& Smithsonian , 60 Garden Street, Cambridge, MA 02138, USA}



\begin{abstract}

We have observed the Class I protostar TMC-1A in the Taurus molecular cloud using the Submillimeter Array (SMA) and the Atacama Large Millimeter/submillimeter Array (ALMA) in the linearly polarized 1.3 mm continuum emission at angular resolutions of $\sim 3\arcsec$ and $\sim 0\farcs 3$, respectively. The ALMA observations also include CO, $^{13}$CO, and C$^{18}$O $J=2-1$ spectral lines.
The SMA observations trace magnetic fields on the 1000-au scale, the directions of which are neither parallel nor perpendicular to the outflow direction. Applying the Davis-Chandrasekhar-Fermi method to the SMA polarization angle dispersion, we estimate a field strength in the TMC-1A envelope of 1-5 mG. It is consistent with the field strength needed to reduce the radial infall velocity to the observed value, which is substantially less than the local free-fall velocity. The ALMA polarization observations consist of two distinct components -- a central component and a north/south component. The central component shows polarization directions in the disk minor axis to be azimuthal, suggesting dust self-scattering in the TMC-1A disk. The north/south component is located along the outflow axis and the polarization directions are aligned with the outflow direction. We discuss possible origins of this polarization structure, including grain alignment by a toroidal magnetic field and mechanical alignment by the gaseous outflow.
In addition, we discover a spiral-like residual in the total intensity (Stokes $I$) for the first time. The C$^{18}$O emission suggests that material in the spiral-like structure is infalling at a speed that is 20\% of the local Keplerian speed.

\end{abstract}


\section{Introduction} \label{sec:intro}

Magnetic fields are expected to play an important role in the process of star formation. During the main accretion phase, magnetic fields in protostellar envelopes regulate mass and momentum accretion onto a disk and consequently affect disk formation and disk properties, such as mass and size. Thus, magnetic fields may have an impact on the subsequent planet formation in the system, since such protostellar disks provide the initial conditions for planet formation. On one hand, magnetohydrodynamics (MHD) simulations have suggested that magnetic fields strongly suppress disk formation in the protostellar phase through the so-called the magnetic braking \citep[e.g., ][]{me.li08, he.ci09, joos12}. On the other hand, theoretical studies also suggest that non-ideal MHD effects (Ohmic dissipation, ambipolar diffusion, and Hall effect) can weaken the effects of magnetic braking, enhancing disk formation \citep[e.g., ][]{inut10, kras11, dapp12, tomi15, tsuk15, zhao18}. Observations of magnetic fields are thus necessary to verify the above theoretical predictions and reveal how magnetic fields affect mass accretion onto disks in the disk formation phase.

Polarized emission from dust grains enables us to observe magnetic fields. However, recent observational studies have reported that magnetic fields are not the only source of polarized emission on disk scales in young stellar objects (YSOs). For example, protoplanetary disks show polarized dust emission arising from self-scattering \citep{kata16, kata17, ohas19} at millimeter wavelengths. This interpretation is based on theoretical predictions of the polarization direction and fraction: polarization directions along the disk minor axis and polarization fractions on the order of 1\% independent of the polarization intensity \citep{yang17}. In addition to self-scattering, the mechanical momentum in the protostellar outflow can align dust grains producing polarized dust emission. If gaseous flux aligns the major axis of dust grains along the outflow direction \citep{gold52}, the resultant polarization direction is parallel to the outflow axis. In contrast, if torque generated by the outflow rotates dust grains around the outflow axis, the resultant polarization direction is perpendicular to the outflow axis \citep{lego19}.

Recent observations toward protostars on multiple spatial scales suggest that magnetic fields affect circumstellar materials in different degrees on different spacial scales \citep{hull17a, hull17b}. It is, therefore, important to observationally investigate magnetic fields, along with kinematics, specifically on the protostellar envelope-disk scales in order to understand how magnetic fields regulate mass accretion onto disks.



In order to investigate the effects of magnetic fields on mass accretion onto a protostellar disk, we observed the Class I protostar TMC-1A using the SMA and ALMA with the polarized dust continuum emission at 1.3 mm. TMC-1A is a good target for this purpose because previous observational studies have already revealed the kinematics in its protostellar disk and envelope. TMC-1A is in the Taurus molecular cloud located at a distance of 130-160 pc away from us \citep{gall18}; we adopt 140 pc as the distance to TMC-1A in this paper. The disk in TMC-1A was identified kinematically by a power-law index of radial profiles of rotational velocity close to $-0.5$ \citep{hars14, aso15}. \citet{hars14} fitted a disk and envelope model to the continuum visibilites at 1.3 mm derived from observations using Plateau de Bure Interferometer (PdBI) at an angular resolution of $0\farcs 7$-$1\farcs 3$, estimating the disk size, disk inclination angle, and central stellar mass to be 80-100 au, $55\arcdeg$, and $0.53~\Msun$, respectively. \citet{aso15} reproduced, by a disk and envelope model, position-velocity (PV) diagrams along the major- and minor-axes of the disk in the C$^{18}$O $J=2-1$ line observed with ALMA at an angular resolution of $\sim 1\arcsec$, estimating the disk size, disk inclination angle, and central stellar mass to be $\sim 100$ au, $\sim 65\arcdeg$, and $0.68~\Msun$, respectively. They also suggested an radial infall velocity of $\sim 0.3$ times the free fall velocity from their model and estimated a magnetic-field strength required to explain the slow infall velocity to be $\sim 2$~mG. In addition to the disk and envelope, \citet{bjer16} revealed disk wind from a radius range of 7-22 au from the disk in TMC-1A. Previous observations at a 8-au resolution revealed that optically thick dust hides the C$^{18}$O and $^{13}$CO $J=2-1$ lines in the central $\pm40$ au region \citep{hars18}. The authors interpreted this high optical depth as a result of dust grains with mm size rather than a result of a massive disk because the high mass required to explain the high optical depth should cause gravitational instability, which is inconsistent with the smooth structure observed in the continuum emission.

This paper is structured as follows. We describe the settings of our SMA and ALMA observations in Section \ref{sec:obs}. Section \ref{sec:results} shows results of polarized 1.3 mm coninuum emission observed with the SMA and ALMA, and Stokes $I$ of the CO, $^{13}$CO, and C$^{18}$O $J=2-1$ lines observed with ALMA. We apply the Davis-Chandrasekhar-Fermi (DCF) method to the SMA result and analyze a non-axisymmetric component in the ALMA continuum image in Section \ref{sec:analyses}. Possible origins of the observed polarization and the non-axisymmetric component are discussed in Section \ref{sec:discussion}. We summarize our conclusions in Section \ref{sec:conclusions}.

\clearpage

\section{SMA and ALMA Observations} \label{sec:obs}
\subsection{SMA} \label{sec:sma}
We observed TMC-1A using the SMA for one track on 2018 January 3 with the full polarization mode. The total observing time is $\sim525$ min (8.75 hr) for TMC-1A including overhead. Seven antennas were used in the compact configuration. The minimum projected baseline length is 16 m. Any emission beyond $9\arcsec$ (1300 au) was resolved out by $\gtrsim 63\%$ with the antenna configuration \citep{wi.we94}. Our SMA observations used two orthogonally polarized receivers, tuned to the same frequency range in the full polarization mode, and the SWARM correlator. The upper sideband (USB) and lower sideband (LSB) covered 213--221 and 229--237 GHz, respectively. Each sideband was divided into four `chunks' with a bandwidth of 2 GHz, and each 'chunk' has a fixed channel width of 140 kHz.

The SMA data were calibrated with the MIR software package\footnote{https://github.com/qi-molecules/sma-mir}. Instrumental polarization leakage was calibrated independently for USB and LSB using the MIRIAD task {\it gpcal} and removed from the data. 
The calibrated visibility data in Stokes $I$, $Q$, and $U$ were Fourier transformed and CLEANed with the natural weighting, using the MIRIAD package \citep{saul95}, by integrating channels without line emission. The polarized intensity, position angle, and polarization percentage were derived from the Stokes $I$, $Q$, and $U$ maps using the MIRIAD task {\it impol}. The parameters of our SMA observations mentioned above and others are summarized in Table \ref{tab:sma}.

\begin{deluxetable*}{cc}
\tablecaption{Summary of the SMA observational parameters \label{tab:sma}}
\tablehead{
\colhead{Date} & 2018.January.03\\
\colhead{Projected baseline length} & \colhead{16--77 m (12--58 k$\lambda$)}\\
\colhead{Primary beam} & \colhead{$54\arcsec$}\\
\colhead{Passband/polarization calibrator} & \colhead{3C279 (7.2 Jy), 3C454.3 (12 Jy)}\\
\colhead{Flux calibrator} & \colhead{Neptune}\\
\colhead{Gain/polarization calibrator} & \colhead{0510+180 (1.5 Jy), NRAO530 (1.5 Jy)}\\
\colhead{Coordinate center (J2000)} & \colhead{$04^{\rm h}39^{\rm m}$35\fs 20, $25^{\circ}41\arcmin 44\farcs 35$}
}
\startdata
Frequency (GHz) & 224 \\
Bandwidth (GHz) & 8 \\
Beam (P.A.) & $2\farcs 8\times 2\farcs4\ (47\arcdeg)$ \\
rms noise level ($\mJB$) & 0.75 (Stokes I), 0.26 (Stokes Q/U)
\enddata
\end{deluxetable*}

\subsection{ALMA} \label{sec:alma}
We also observed TMC-1A using ALMA in Cycle 6 on 2018 November 18 with the full polarization mode. The total observing time is $\sim202$ min (3.37 hr), while the on-source observing time for TMC-1A is $\sim 43$ min. The number of antennas was 49. The array configuration was C43-5, whose minimum projected baseline length is 12 m. Any emission beyond $12\arcsec$ (1700 au) was resolved out by $\gtrsim 63\%$ with the antenna configuration \citep{wi.we94}. Spectral windows for $^{12}$CO, $^{13}$CO, and C$^{18}$O ($J=2-1$) lines have 1920, 960, and 960 channels covering 117, 59, and 59 MHz band widths at frequency resolutions of 61.0, 61.0, and 61.0 kHz, respectively. In making maps, channels are binned to produce the velocity resolution of $0.1\ \kms$ for all of the three lines. Another spectral window covers 232-234 GHz, which was assigned to the continuum emission.

All the imaging procedure was carried out with Common Astronomical Software Applications (CASA), version 5.6.2. The visibilities in Stokes $I$, $Q$, $U$, and $V$ were Fourier transformed and CLEANed with Briggs weighting, a robust parameter of 0.5, and a threshold of 3$\sigma$. Emission of the three lines were detected only in Stokes $I$. The CLEAN process using the CASA task {\it tclean} adopted the auto-masking (auto-multithresh) with the parameters of sidelobethreshold=3.0, noisethreshold=3.0, lownoisethreshold=1.5, and negativethreshold=3.0. The polarized intensity and position angle were derived from the Stokes $I$, $Q$, and $U$ maps using the CASA task {\it immath}.

We also performed self-calibration for Stokes $I$ of the continuum data using tasks in CASA ($clean$, $gaincal$, and $applycal$).
Only the phase was calibrated by decreasing the time bin from `inf', 90 sec, and then 18 sec. The self-calibration improved the rms noise level of the continuum maps by a factor of $\sim 120$. The obtained calibration tables for the Stokes $I$ continuum data were also applied to the other Stokes continuum data and the line data for consistency. The noise level of the line maps were measured in emission-free channels. The parameters of our ALMA observations mentioned above and others are summarized in Table \ref{tab:alma}.

\begin{deluxetable*}{ccccc}
\tablecaption{Summary of the ALMA observational parameters \label{tab:alma}}
\tablehead{
\multicolumn{2}{c}{Date}&\multicolumn{3}{c}{2018.November.21}\\
\multicolumn{2}{c}{Projected baseline length}&\multicolumn{3}{c}{12--1376 m (9.4--1070 k$\lambda$)}\\
\multicolumn{2}{c}{Primary beam}&\multicolumn{3}{c}{$27\arcsec$}\\
\multicolumn{2}{c}{Bandpass/flux calibrator}&\multicolumn{3}{c}{J0510+1800}\\
\multicolumn{2}{c}{Polarization calibrator}&\multicolumn{3}{c}{J0522-3627 (1.0 Jy)}\\
\multicolumn{2}{c}{Check source}&\multicolumn{3}{c}{J0426+2952}\\
\multicolumn{2}{c}{Phase calibrator}&\multicolumn{3}{c}{J0438+3004 (0.23 Jy)}\\
\multicolumn{2}{c}{Coordinate center (ICRS)}&\multicolumn{3}{c}{$04^{\rm h}39^{\rm m}$35\fs 20, $25^{\circ}41\arcmin 44\farcs 23$}
}
\startdata
&Continuum&CO $J=2-1$&$^{13}$CO $J=2-1$&C$^{18}$O $J=2-1$\\
\hline
Frequency (GHz)&234&230.538000&220.398684&219.560354\\
Bandwidth/resolution&2 GHz&$0.1\ \kms$&$0.1\ \kms$&$0.1\ \kms$\\
Beam (P.A.)&$0\farcs 37\times 0\farcs26\ (-2\arcdeg)$&$0\farcs 40\times 0\farcs 28\ (-1\arcdeg)$&$0\farcs 42\times 0\farcs 29\ (-1\arcdeg)$&$0\farcs 42\times 0\farcs 29\ (-1\arcdeg)$\\
rms noise level&$25~\uJB$&$2~\mJB$&$2~\mJB$ & $2~\mJB$
\enddata
\end{deluxetable*}

\clearpage

\section{Results} \label{sec:results}
\subsection{SMA Polarized 1.3 mm Continuum}

\begin{figure*}[htbp]
\gridline{
\fig{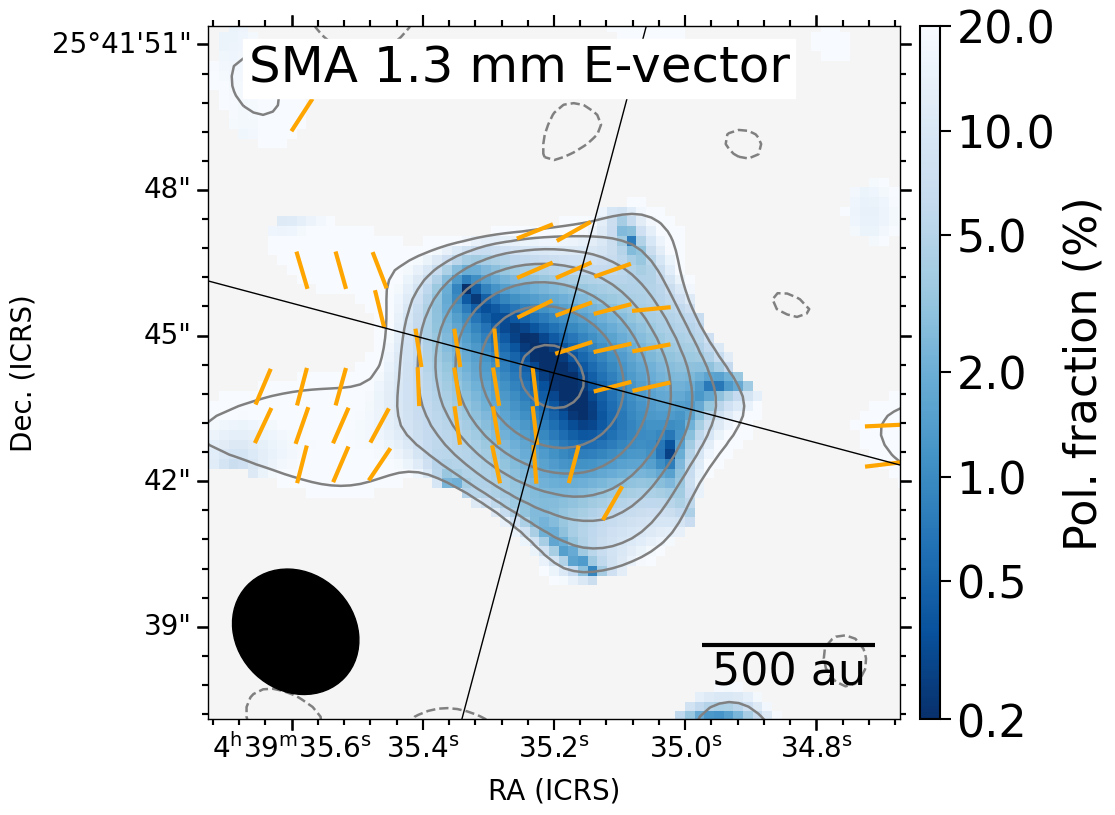}{0.36\textwidth}{(a)}
\fig{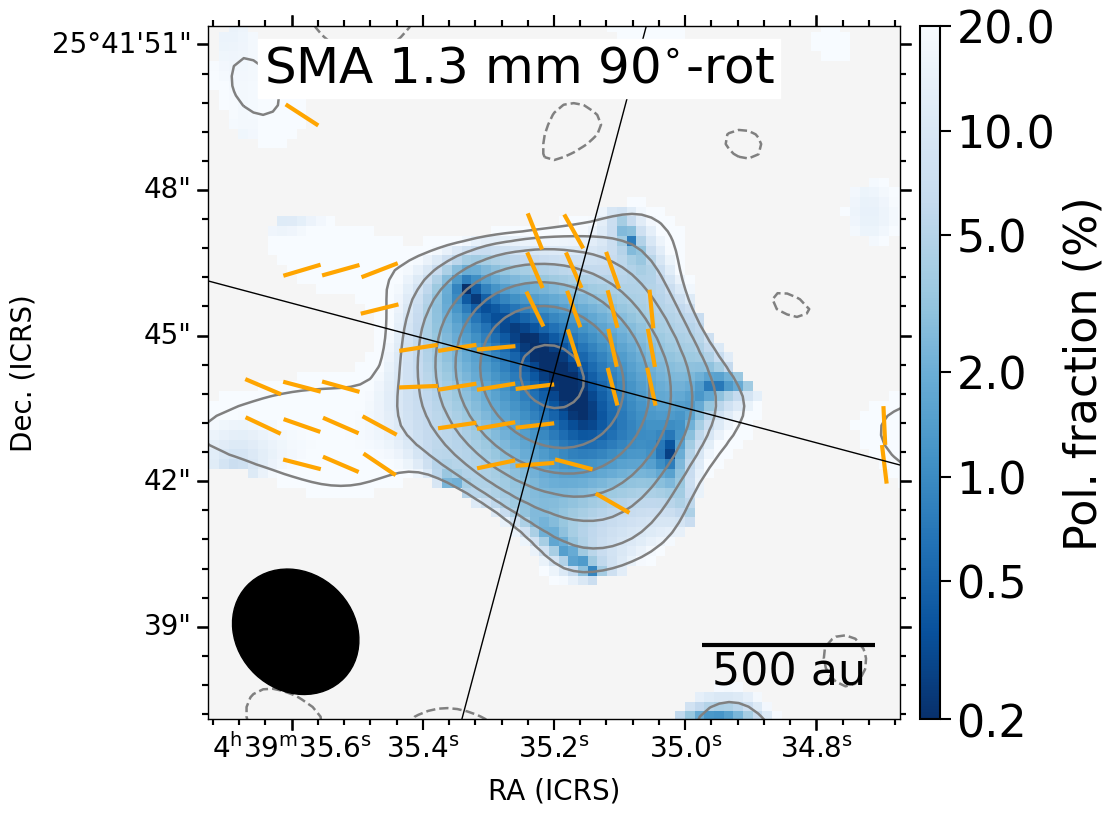}{0.36\textwidth}{(b)}
}
\caption{The continuum emission at 1.3 mm observed with the SMA. The contour map shows Stokes $I$ with contour levels of 3, 6, 12, 24,...$\times \sigma$, where $1\sigma$ corresponds to $0.75\ \mJB$. The color map shows polarization fraction, not de-biased, in the region where Stokes $I$ is above the $2\sigma$ level. The segments show (a) polarization angles and (b) those rotated by $90\arcdeg$, de-biased with the 2$\sigma$ ($0.52\ \mJB$) cutoff. The diagonal lines denote the major (P.A.$=75\arcdeg$) and minor ($165\arcdeg$) axes of the TMC-1A disk (Section \ref{sec:sym}), centered at the protostellar position. The filled ellipse denotes the SMA synthesized beam.
\label{fig:smacont}}
\end{figure*}

Figure \ref{fig:smacont} shows the polarized continuum emission at 1.3 mm observed with the SMA. The Stokes I map shows a compact component associated with TMC-1A and an extension to the east at the $3\sigma$ level, and slightly expands into the southwest. Gaussian fitting to the Stokes I emission provides a peak position of $\alpha ({\rm J2000})=04^{\rm h}39^{\rm m}35\fs 20,\ \delta ({\rm J2000})=25\arcdeg 41\arcmin 44\farcs 14$ and a deconvolved size of $0\farcs 66 \pm 0\farcs 09\times 0\farcs 50 \pm 0\farcs 11$ with a position angle of $66\arcdeg \pm 28\arcdeg$. This size corresponds to $92\times 70$~au at the distance of TMC-1A, 140 au. The major axis of the emission is consistent with the major axis of the disk around TMC-1A and perpendicular to the associated outflow \citep{aso15}. The peak intensity and total flux density measured in our SMA observation are $0.171~\JB$ and 0.182~Jy, respectively.

The polarized fraction $\sqrt{Q^2 + U^2} / I$, where $I$, $Q$, and $U$ are the Stokes parameters, is shown in the region where Stokes $I$ is detected above the $2\sigma$ level.
The orange segments have a uniform length and show the polarization angles in panel (a) and those rotated by $90\arcdeg$ in panel (b) that are de-biased with the $2\sigma$ level. The polarization fraction is typically $\sim 10\%$ in the north and east of TMC-1A, where the de-biased polarization is detected. In contrast, the fraction is $\lesssim 0.5\%$, i.e., de-polarized in the northeast, center, and southwest. The polarization angle is $\sim -45\arcdeg$ from the disk-minor axis (Section \ref{sec:sym}) on the northern side of the de-polarized layer, while being distributed around $\sim +30\arcdeg$ from the disk minor axis on the southern side of the de-polarized layer. The angle is around the disk minor axis in the eastern extension.

The $90\arcdeg$-rotated segments (Figure \ref{fig:smacont}b) are supposed to trace the direction of magnetic fields in the protostar on the 1000-au scale observed with the SMA. The $90\arcdeg$-rotated segments have relative angles to the disk minor axis (P.A.$\sim -15\arcdeg$; see Section \ref{sec:sym}) ranging from $\sim 20\arcdeg$ to $\sim 50\arcdeg$ in the northwest, while the relative angles range from $\sim -80\arcdeg$ to $\sim -60\arcdeg$ in the southeast. The inferred magnetic field morphology shows symmetry with respect to the de-polarized layer. The inferred field appears to be drawn from the northeast to the southwest turning the direction at the de-polarized layer. When the variation of the field is smaller than an observational beam size, the region appears to be de-polarized. Hence, drawn morphology of the magnetic fields may explain the observed de-polarized layer. The de-polarization due to such field variation is numerically simulated \citep{kata12} and observed in other protostellar systems \citep[e.g.,][]{kwon19}.


\subsection{ALMA Polarized 1.3 mm Continuum} \label{sec:almacont}

\begin{figure*}[htbp]
\gridline{
\fig{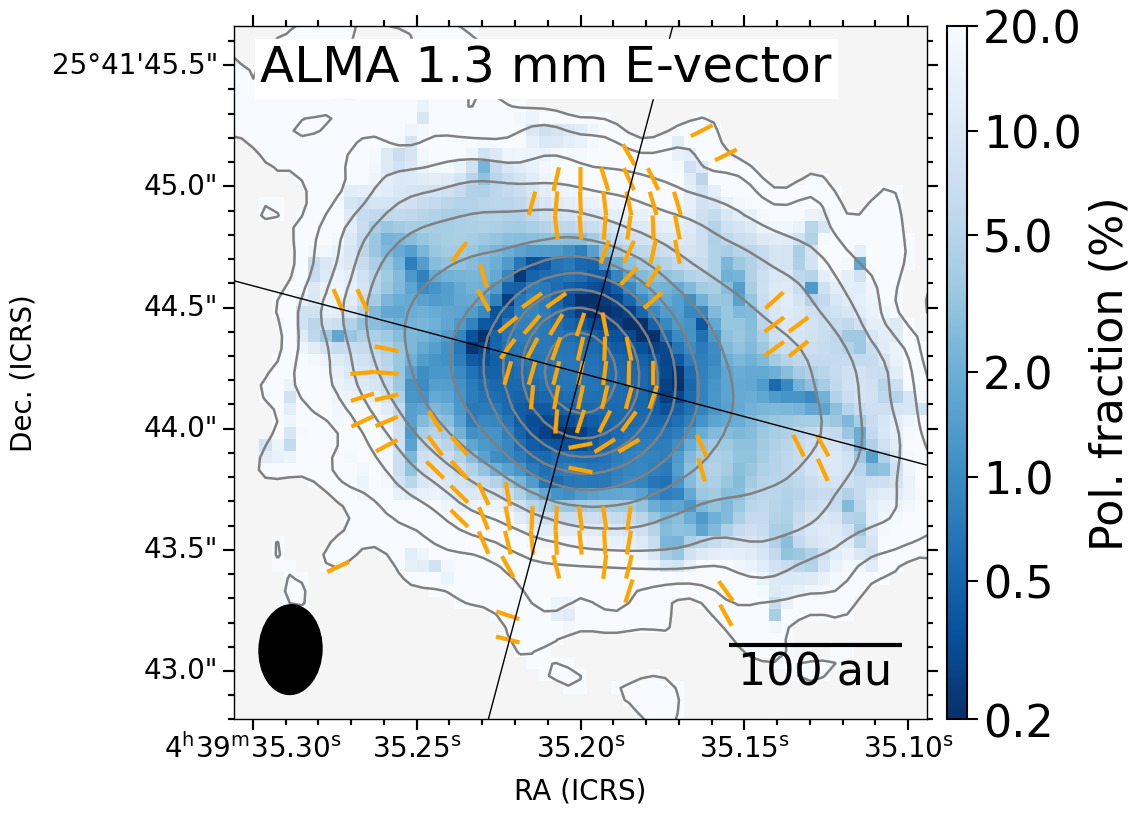}{0.36\textwidth}{(a)}
\fig{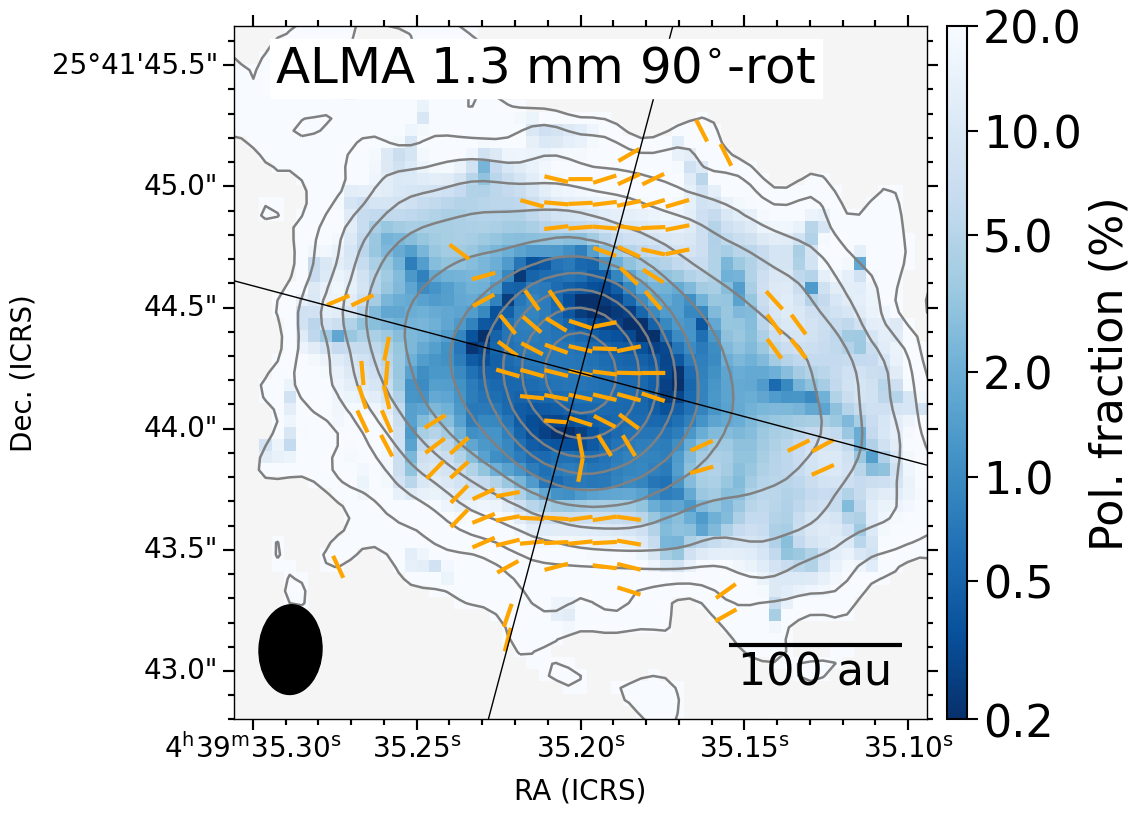}{0.36\textwidth}{(b)}
}
\caption{The continuum emission at 1.3 mm observed with ALMA. The contour map shows Stokes $I$ with contour levels of 3, 6, 12, 24,...$\times \sigma$, where $1\sigma$ corresponds to $25\ \uJB$. The color map shows polarization fraction, not de-biased, in the region where Stokes $I$ is above the $3\sigma$ level. The segments show (a) polarization angles and (b) those rotated by $90\arcdeg$, de-biased with the 3$\sigma$ ($75\ \uJB$) cutoff. The diagonal lines denote the major (P.A.$=75\arcdeg$) and minor ($165\arcdeg$) axes of the TMC-1A disk (Section \ref{sec:sym}), centered at the protostellar position. The filled ellipse denotes the ALMA synthesized beam.
\label{fig:almacont}}
\end{figure*}

Figure \ref{fig:almacont} shows the polarization continuum emission at 1.3 mm observed with ALMA. The spatial scale is five times smaller than the SMA figure, Figure \ref{fig:smacont}. The Stokes $I$ emission consists of a compact, strong component ($\lesssim 50$~au in radius or a signal-to-noise ratio of $\gtrsim 150\sigma$) and an extended component having sizes in radius of $\sim 200$~au in the direction of P.A.$=75\arcdeg$ (the disk major axis in Section \ref{sec:sym}) and $\sim 120$~au in the direction of P.A.$=165\arcdeg$ (the disk minor axis in Section \ref{sec:sym}) at the $3\sigma$ level. The peak intensity and total flux density measured in our ALMA observation are $0.120~\JB$ and 0.236~Jy, respectively. The direction of the major elongation is consistent with previous 1.3 mm observations at a spatial resolution of $\sim 8$~au \citep{bjer16, hars18}. Becuase the Stokes I emission shows the compact and extended components, double-Gaussian fitting is more appropriate for this Stokes $I$ emission than single-Gaussian fitting is. Double-Gaussian fitting provides a peak position of $\alpha ({\rm J2000})=04^{\rm h}39^{\rm m}35\fs 200,\ \delta ({\rm J2000})=25\arcdeg 41\arcmin 44\farcs 229$ for the compact component, while $\alpha ({\rm J2000})=04^{\rm h}39^{\rm m}35\fs 202,\ \delta ({\rm J2000})=25\arcdeg 41\arcmin 44\farcs 235$
for the extended component. We adopt the peak position of the compact component as the protostellar position of TMC-1A in this paper. The deconvolved sizes of the compact and extended components are $0\farcs 25\times 0\farcs 15$ (P.A.$=76\arcdeg$) and $1\farcs 09\times 0\farcs 63$ (P.A.$=73\arcdeg$), corresponding to $35\times 21$~au and $153\times 88$~au, respectively.

\begin{deluxetable*}{ccccc}
\tablecaption{Results of Gaussian fitting to the SMA and ALMA 1.3 mm continuum images. The fitting to the ALMA image adopts a double-Gaussian function. \label{tab:gauss}}
\tablehead{
\colhead{Observations} & \colhead{Flux (mJy)} & 
\colhead{Center}
& \colhead{Deconvolved FWHM} & \colhead{P.A.}
}
\startdata
SMA & $169.8\pm 0.9$ & 
\begin{tabular}{c}
$04^{\rm h}39^{\rm m}35\fs 2024\pm 0.0005{\rm s}$ \\ $25\arcdeg 41\arcmin 44\farcs 145\pm 0.006\arcsec$ 
\end{tabular}
& 
\begin{tabular}{c}
$0\farcs 66\pm 0\farcs 09$ \\ $0\farcs 50 \pm 0\farcs10$ 
\end{tabular}
& $66\arcdeg \pm 28 \arcdeg$ \\
\hline
ALMA compact & $111.66\pm 0.03$ & 
\begin{tabular}{c}
$04^{\rm h}39^{\rm m}35\fs 199945\pm 0.000003{\rm s}$ \\ $25\arcdeg 41\arcmin 44\farcs 22931\pm 0.00005\arcsec$
\end{tabular}
& 
\begin{tabular}{c}
$0\farcs 2502\pm 0\farcs 0002$ \\ $0\farcs 1542 \pm 0\farcs004$ 
\end{tabular}
& $75.8\arcdeg \pm 0.1 \arcdeg$ \\
ALMA extended & $66.8\pm 0.2$ & 
\begin{tabular}{c}
$04^{\rm h}39^{\rm m}35\fs 2017\pm 0.0001{\rm s}$ \\ $25\arcdeg 41\arcmin 44\farcs 235\pm 0.001\arcsec$
\end{tabular}
& 
\begin{tabular}{c}
$1\farcs 085\pm 0\farcs 004$ \\ $0\farcs 632 \pm 0\farcs03$ 
\end{tabular}
& $73.4\arcdeg \pm 0.3 \arcdeg$ \\
\hline
\enddata
\end{deluxetable*}

The polarized fraction is shown in the region where Stokes $I$ is detected above the $3\sigma$ level. The polarization angles are de-biased with the $3\sigma$ level. The polarization fraction is typically $\sim 0.7\%$ at the center, where the de-biased polarization is detected. This central polarized region is surrounded by a de-polarized ring with a radius of $\sim $40-60~au showing a polarization fraction of $\lesssim 0.3\%$. The polarization angle in the central region is overall in the disk minor axis (Section \ref{sec:sym}), whereas it also has the azimuthal component particularly in the outer part of this region.

A similar de-polarized ring is also reported in the disk around a massive protostar, GGD27 MM1 \citep{gira18}. In addition to the ring, the disk around GGD27 MM1 shows spatial distribution of polarization fraction
similar to that of TMC-1A: the fraction is $\sim 1\%$ in an inner part of the disk and $>6\%$ in an outer part of the disk. On the other hand, the inner polarization fraction in GGD27 MM1 is higher on the near side of the disk, from which \citet{gira18} suggests that dust settling has not occurred yet in GGD27 MM1. In comparison, the inner polarization in TMC-1A shows symmetric distribution of the polarization fraction and thus could be interpreted as dust settling stronger than in GGD27 MM1. The outer polarization in GGD27 MM1 shows azimuthal directions, which is ascribed to self-scattering with optically thin continuum emission by the authors. In comparison, the outer polarization direction in TMC-1A is roughly radial, and thus the polarization mechanism in the outer part is unlikely the self-scattering.

In addition to the central polarized component, the de-biased polarization is also detected at $\sim 100$~au north and south of the central protostar, where the polarization fraction is typically $\sim 10\%$. The segments in the northern and southern regions have relative angles to the disk-minor axis (P.A.$= -15\arcdeg$; see Section \ref{sec:sym}) ranging from $\sim 0\arcdeg$ to $\sim 45\arcdeg$ in the northern region, while the relative angles range from $\sim 0\arcdeg$ to $\sim 35\arcdeg$ in the southern region. In other words, the $90\arcdeg$-rotated segments are roughly in the azimuthal direction in the northern and southern regions, as shown in Figure \ref{fig:almacont}(b). The polarization direction in the southern region is more consistent with the SMA result than that in the northern region, although it is difficult to directly compare the SMA and ALMA results because of the spacial scale difference by one order of magnitude. Unlike on the SMA scale, several mechanisms can cause polarization on the 100 au scale around protostars. Potential mechanisms will be discussed in Sections \ref{sec:cpol} and \ref{sec:nspol}.


\subsection{ALMA CO Isotopologue Lines} \label{sec:line}

\begin{figure*}[htbp]
\gridline{
\fig{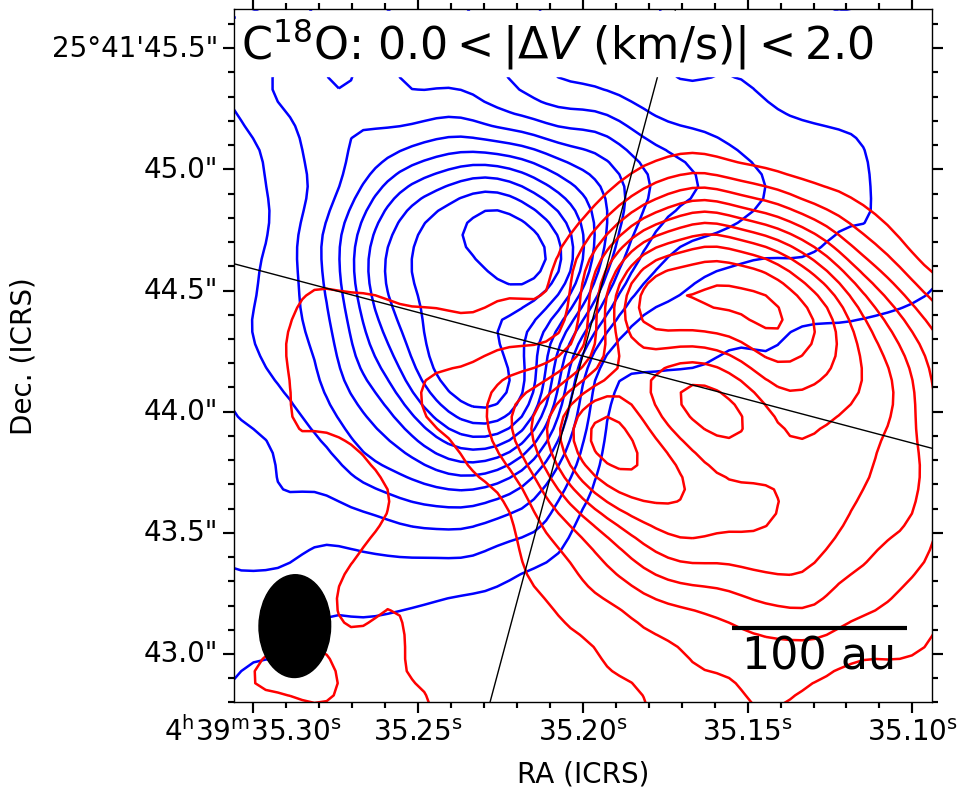}{0.3\textwidth}{(a)}
\fig{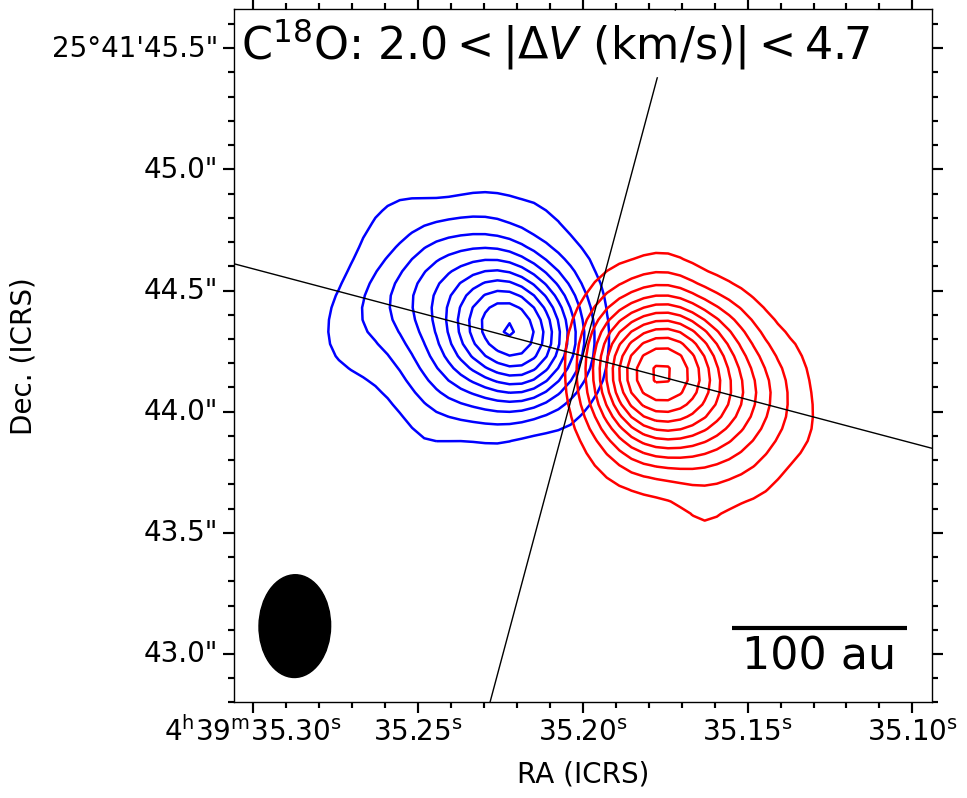}{0.3\textwidth}{(b)}
}
\gridline{
\fig{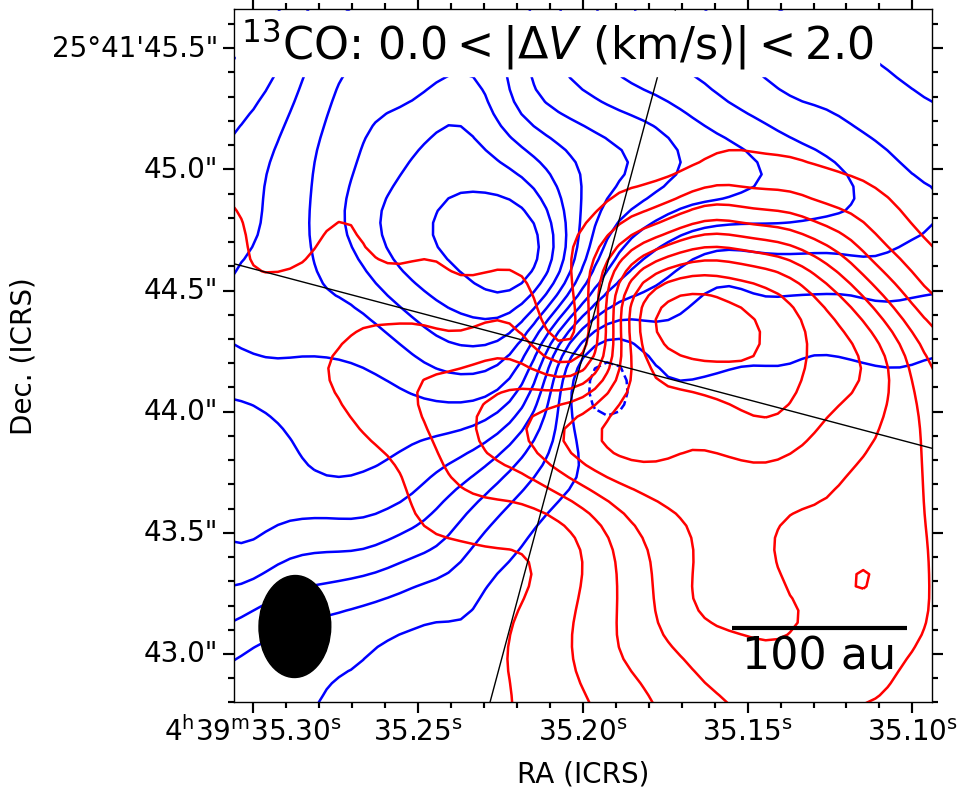}{0.3\textwidth}{(c)}
\fig{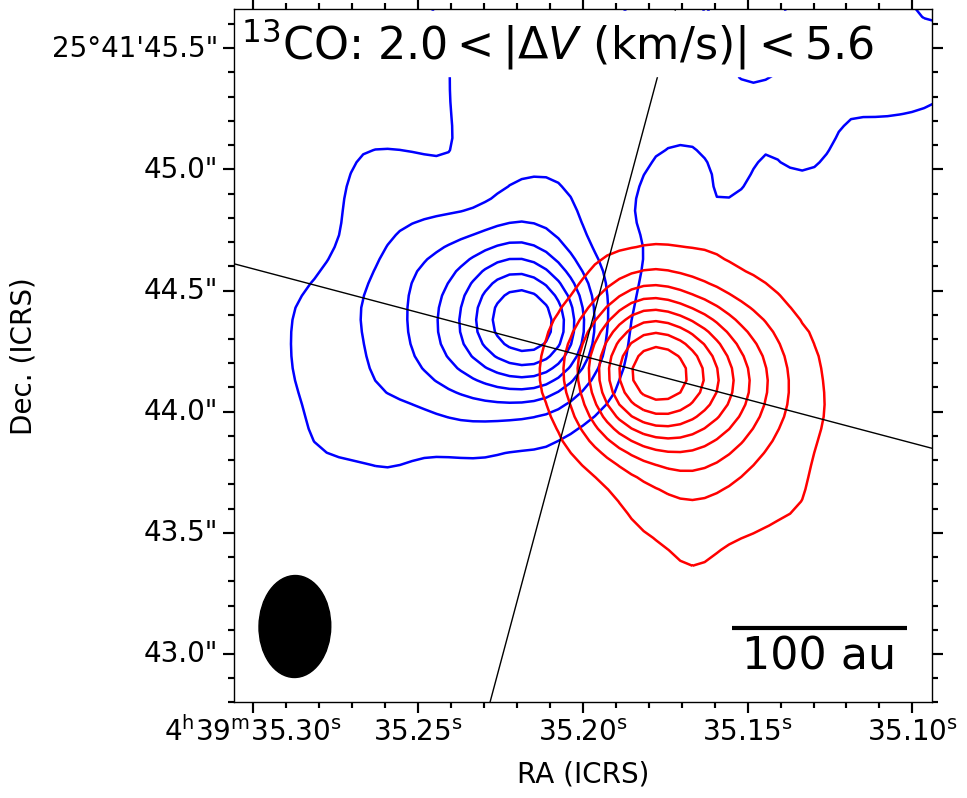}{0.3\textwidth}{(d)}
}
\gridline{
\fig{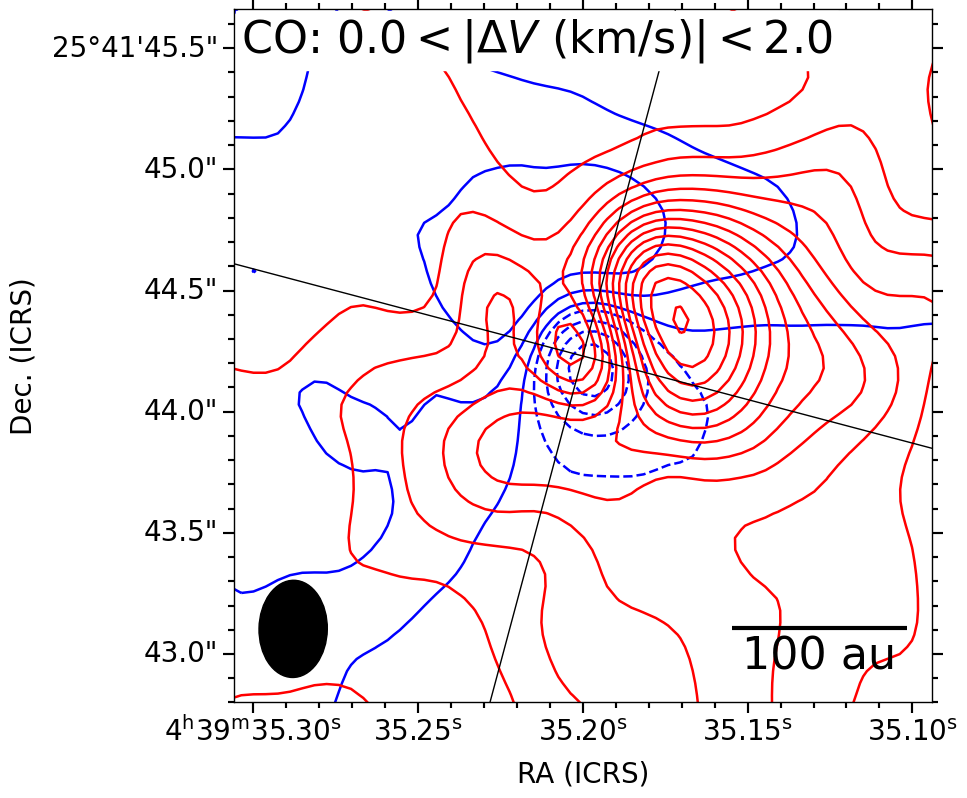}{0.3\textwidth}{(e)}
\fig{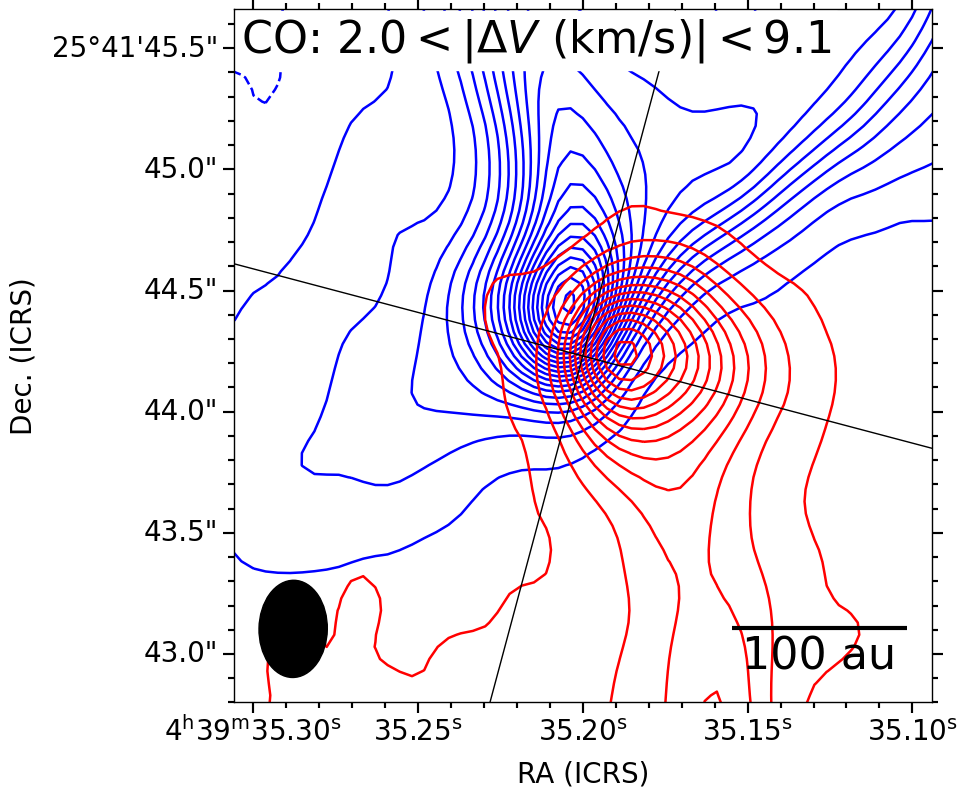}{0.3\textwidth}{(f)}
\fig{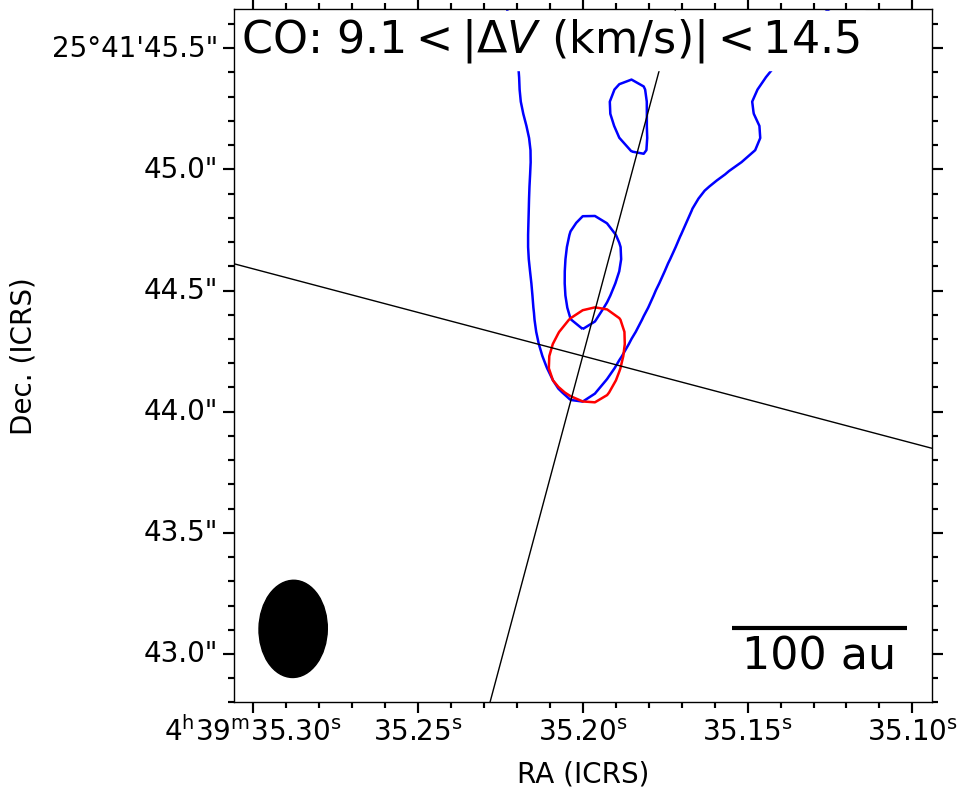}{0.3\textwidth}{(g)}
}
\caption{Integrated channel maps of the CO isotopologue lines observed with ALMA. The integrated velocity range relative to $V_{\rm sys}=6.34\ \kms$ (Section \ref{sec:kep}) is denoted in each panel. Blue and red contours show blue- and redshifted emission, respectively. Contour levels are in $12\sigma$, $24\sigma$, and $36\sigma$ steps for C$^{18}$O, $^{13}$CO and CO, respectively, from $12\sigma$}, where $1\sigma$ corresponds to (a) 0.9, (b) 1.0, (c) 0.9, (d) 1.2, (e) 0.9, (f) 1.7, and (g) 1.5 $\mJB~\kms$. The diagonal lines denote the major (P.A.$=75\arcdeg$) and minor ($165\arcdeg$) axes of the TMC-1A disk (Section \ref{sec:sym}), centered at the protostellar position. The filled ellipse at the bottom-left corner of each panel denotes the ALMA synthesized beam.
\label{fig:line}
\end{figure*}

Figure \ref{fig:line} shows results of the CO, $^{13}$CO, and C$^{18}$O $J=2-1$ lines observed with ALMA. The spatial scale is the same as that of Figure \ref{fig:almacont}. The blue- and redshifted emission is integrated over the same velocity width, from the systemic velocity, and thus has the same noise level in each panel. The integrated range is divided at $2~\kms$, which roughly corresponds to the Keplerian velocity at the disk radius of TMC-1A, $\sim100$~au, with the central protostellar mass, $\sim 0.7~\Msun$, \citep{aso15} and an inclination angle of $50\arcdeg$--$60\arcdeg$. The high-velocity components of the C$^{18}$O and $^{13}$CO emission (Figure \ref{fig:line}b and \ref{fig:line}d) are integrated until the highest velocity at which the emission is detected at the signal-to-noise ratio (SNR) $3\sigma$. The boundary of the high-velocity (Figure \ref{fig:line}f) and very-high-velocity (Figure \ref{fig:line}g) components of the CO emission is the highest velocity at which the blueshifted emission is detected at SNR$=3\sigma$. In these integrated channel maps on the same scale as the continuum image, the entire emission is shown in Appendix \ref{sec:mom01} using the integrated intensity (moment 0) and mean velocity (moment 1) maps.

The C$^{18}$O emission shows a velocity gradient mainly along the disk major axis (Section \ref{sec:sym}) in both the low- (Figure \ref{fig:line}a) and high-velocity (Figure \ref{fig:line}b) components. In addition to this main velocity gradient, the low-velocity component also shows blueshifted emission in the northwest and redshifted emission in the southeast at SNR $\lesssim 50\sigma$, causing a different, diagonal velocity gradient. This diagonal velocity gradient is reported in \citet{aso15} in the low velocity range $<2~\kms$. The low-velocity component of the C$^{18}$O emission shows double-peaked morphology on both blue- and redshifted sides, whereas the high-velocity component shows compact, single-peaked morphology on both blue- and redshifted sides. The double-peaked morphology in the low velocity range is newly revealed with the higher angular resolution than in \citet{aso15}.

The $^{13}$CO emission shows the same features as the C$^{18}$O emission: the main velocity gradient along the disk major axis, another diagonal velocity gradient in the low-velocity component at SNR $\lesssim 120\sigma$, and compact, single-peaked morphology in the high-velocity component. 
The emission peaks make the main velocity gradient along the disk major axis in both high- and low-velocity components, although the emission peaks are shifted to the north from the disk major axis in the low-velocity component. Emission at low SNR makes the diagonal velocity gradient shown in the low-velocity C$^{18}$O emission above and that in \citet{aso15}.
The low-velocity emission decreases in the central $\sim 40$~au, i.e., within one beam. This is consistent with the absorption in the $^{13}$CO $J=2-1$ line due to strong continuum emission at the protostellar position, reported with the observation at a $\sim 8$~au resolution \citep{hars18}.

The CO emission shows more complicated structures than the C$^{18}$O and $^{13}$CO emission. The fainter, or absorbed, part in the low-velocity component is clearer in the CO line than in the $^{13}$CO line. The high-velocity component (Figure \ref{fig:line}f) clearly traces the associated outflow going along the disk minor axis. Strong emission (SNR$>150\sigma$) shows the main velocity gradient along the disk major axis same as the C$^{18}$O and $^{13}$CO emission. A part of the strong emission is also extended to the north tracing the outflow, as is also seen in the $^{13}$CO emission. Weak emission, SNR$<150\sigma$, is extended to the southern side along the disk minor axis in both blue- and redshifted components in the high velocity range (Figure \ref{fig:line}f); these structures are not seen in the C$^{18}$O or $^{13}$CO lines. The very-high-velocity component (Figure \ref{fig:line}g) traces a part of the blueshifted outflow lobe, being more collimated than the high-velocity component (Figure \ref{fig:line}f). This is consistent with the previous observation showing that emission in the blueshifted lobe is more collimated at higher velocities \citep{aso15}.

\clearpage

\section{Analyses} \label{sec:analyses}
\subsection{DCF Method} \label{sec:dcf}

TMC-1A has an infalling rotating envelope around the Keplerian disk. \citet{aso15} reported a radial infall velocity of $\sim 0.3$ times the free-fall velocity and suggested that a magnetic field of $\sim 2$~mG in strength is required to reduce the infall velocity to the observed value. Our SMA observations enable us to test this suggestion by estimating the field strength. The Davis-Chandrasekhar-Fermi (DCF) method \citep{davi51, ch.fe53} is the most widely used technique for inferring the magnetic field strength from polarization observations \citep[e.g., ][]{kwon19}. The method assumes that Alfv\'enic fluctuation dominates the magnetic and velocity fields, with the magnetic-field strength $B_{\rm pos}$ in the plane-of-sky estimated from: 
\begin{equation} \label{eq:dcf}
    B_{\rm pos} = \xi \sqrt{4\pi \langle \rho \rangle} \frac{\delta v_{\rm los}}{\delta \phi},
\end{equation}
where $\langle \rho \rangle$, $\delta v_{\rm los}$, and $\delta \phi$ are the mean density, dispersion of the line-of-sight velocity, and dispersion of the polarization angle, respectively. The correction factor $\xi$ is usually taken to be $\sim 0.5$ based on numerical simulations \citep{heit01, ostr01, pado01}.

\begin{figure*}[htbp]
\gridline{
\fig{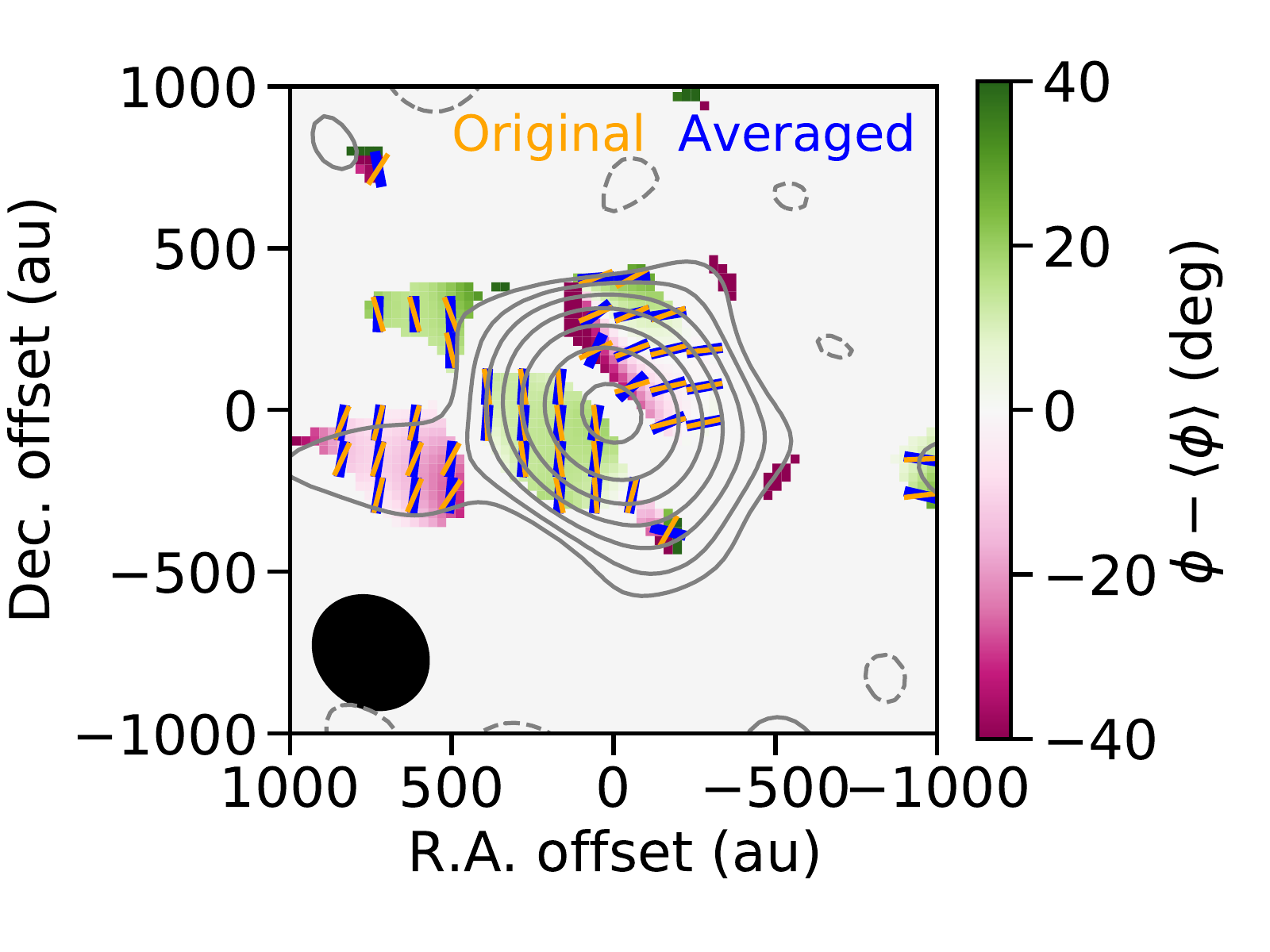}{0.42\textwidth}{(a)}
\fig{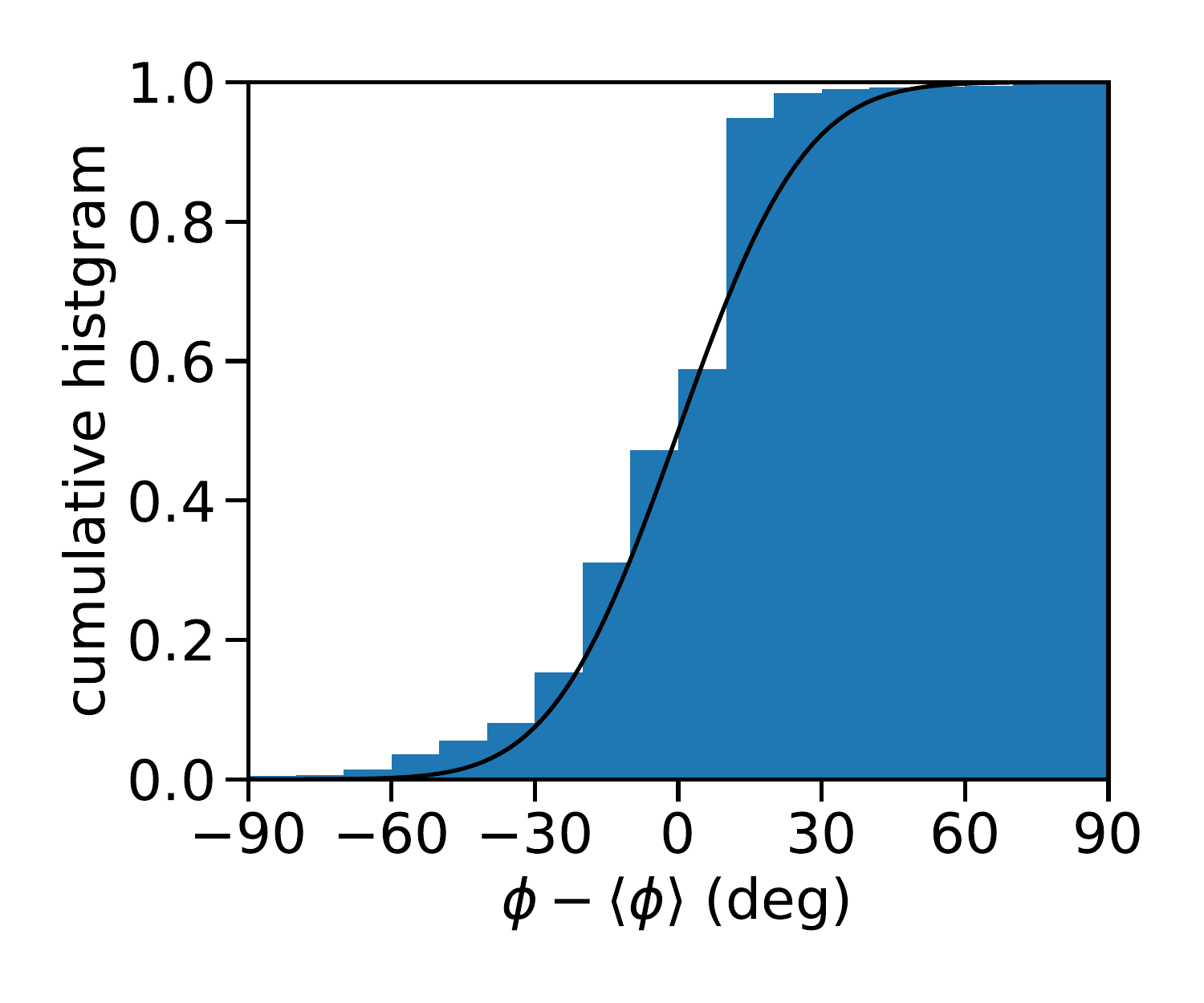}{0.37\textwidth}{(b)}
}
\caption{(a) Deviation of polarization angles observed with the SMA from 2-beam averaged angles. The blue segments denote the 2-beam averaged angles $\langle \phi \rangle$. The orange segments, the contour map, and the filled ellipse are the same as those in Figure \ref{fig:smacont}. The coordinates are relative to the protostellar position. (b) Cumulative histogram of the relative polarization angle from the 2-beam averaged angle, $\phi - \langle \phi \rangle$. The black curve is the error function with a standard deviation of $21\arcdeg$.
\label{fig:dcf}}
\end{figure*}

Measuring the angle dispersion $\delta \phi$ requires an average angle at each position. We define the average angle $\langle \phi \rangle$ by averaging the Stokes Q and U emission over a 2D Gaussian function with a FWHM of $6\arcsec$. This Gaussian function is larger than the SMA synthesized beam by a factor of $\sim 2$. This size is selected because the polarized emission is detected over $\sim 4$ beam area. Figure \ref{fig:dcf}(a) shows the average angles as blue segments and the relative angle ($\phi - \langle \phi \rangle$) as a color map. The dispersion $\delta \phi$ in Equation (\ref{eq:dcf}) is derived as the standard deviation of the relative angle, $21\arcdeg$. Figure \ref{fig:dcf}(b) shows the cumulative histogram of the relative angle overlaid with the error function with a standard deviation of $21\arcdeg$. This histogram indicates that the standard deviation represents the distribution of the relative angle well.

The mean density $\langle \rho \rangle$ is derived from the total flux density and the size of the SMA continuum emission. The total flux density 0.182~Jy corresponds to $\sim 0.024~\Msun$ for a dust opacity coefficient, index, dust temperature, and gas-to-dust mass ratio of $0.035~{\rm cm}^2~{\rm g}^{-1}$ at $850~\micron$ \citep{andr05}, $1.46$, 28~K \citep{chan98}, and 100, respectively. The SMA continuum emission is detected over $\sim 500$~au in radius at the $3\sigma$ level. Averaging the mass of $0.024~\Msun$ over a sphere of 500~au in radius, we obtain a mean density of $\sim 3\times 10^{-17}~{\rm g}~{\rm cm}^{-3}$, which is adopted as the mean density $\langle \rho \rangle$ in Equation (\ref{eq:dcf}).

The velocity dispersion is difficult to measure directly from the observations toward TMC-1A because this protostar is known to have ordered motions, such as rotation and/or radial infall, which are not included in the original DCF method. For this reason, the observed velocity dispersion provides an upper limit of $\delta v$ in Equation (\ref{eq:dcf}). For example, the C$^{18}$O emission observed in our ALMA observations has a velocity dispersion (standard deviation) of $\sim 1.0~\kms$ when the emission over the entire spatial area is included.

With the three quantities derived above, as well as the correction factor 0.5, Equation (\ref{eq:dcf}) yields $B_{\rm pol}\sim 3$~mG. This estimate is consistent with the prediction in \citet{aso15} for explaining the radial infall velocity of $\sim 0.3$ times the free fall velocity in the TMC-1A envelope. We caution the reader that the estimate is very rough because each quantity has a large uncertainty. For example, if the average area for $\delta \phi$ is changed from 1.5- to 4-beams, $\delta \phi^{-1}$ would vary by $\pm 50\%$. If dust grains grow to such a large size that the dust opacity index is close to zero, $\sqrt{\langle \rho \rangle}$ would decrease by $\sim 40\%$. If the velocity dispersion due to Alfv\'enic motions is similar to the sound speed at 28 K \citep{chan98}, $\sim 0.3~\kms$, $\delta v$ would decrease by a factor of $\sim 3$. When these uncertainties are taken into account, $B_{\rm los}$ is estimated to be within a range of 1--5~mG.


\subsection{Axisymmetric Component Subtraction} \label{sec:sym}

The morphology of continuum emission helps us reveal the physical origin of the dust polarization. For this purpose, we investigate the morphology of the 1.3 mm continuum emission in our ALMA observations in this subsection. Specifically, we decompose the continuum emission into an axisymmetric component and a residual through model fitting.

Our axisymmetric model has a radial intensity profile, $f(r)$, centered at $(x_0, y_0)$. The central coordinates are free parameters, defined as offsets from the protostellar position derived from the double-Gaussian fitting in Section \ref{sec:almacont}. This model is scaled along the minor axis direction at $\theta _0-90\arcdeg$ by a factor of $\cos i$ and then convolved with the Gaussian beam with the same FWHM as the ALMA synthesized beam. The position angle $\theta _0$ and the inclination angle $i$ are free parameters. To produce an arbitrary radial profile, $f(0~{\rm au})$, $f(20~{\rm au})$, $f(40~{\rm au})$, ..., $f(300~{\rm au})$ are free parameters in this model fitting. The grid spacing of $f(r)$, 20~au, is half beam of the ALMA observations. Intensities at locations with 300~au but not on the grid radii are interpolated from the grid points, and those beyond $300~{\rm au}$ are set to zero. The best-fit parameters are derived by minimizing $\chi ^2=\sum (f_{\rm obs}-f_{\rm mod})^2/\sigma ^2$ in the image domain, where $f_{\rm obs}$, $f_{\rm mod}$, and $\sigma$ are the observed intensity, model intensity, and rms noise level of the continuum observations, respectively. 
For this fitting, we adopt the Markov Chain Monte Carlo method using the open code {\it ptemcee} \citep{fore13, vous16}, where the log-likelihood function is $-\chi ^2 / 2$. The numbers of temperatures, walkers, and the number of steps are 2, 190, and 10000, respectively. The first 5000 steps are burnt out. The uncertainties of the parameters are defined as the 10th and 90th percentiles of the parameter chain.
The best-fit coordinates and angles, including their uncertainties, are $(x_0,\ y_0,\ \theta _0,\ i)=(0.08^{+0.02}_{-0.02}~{\rm au},\ 0.02^{+0.02}_{-0.02}~{\rm au},\ 75^{+0.3}_{-0.2}~{\rm deg},\ 53^{+0.2}_{-0.3}~{\rm deg})$. 

\begin{figure*}[htbp]
\gridline{
\fig{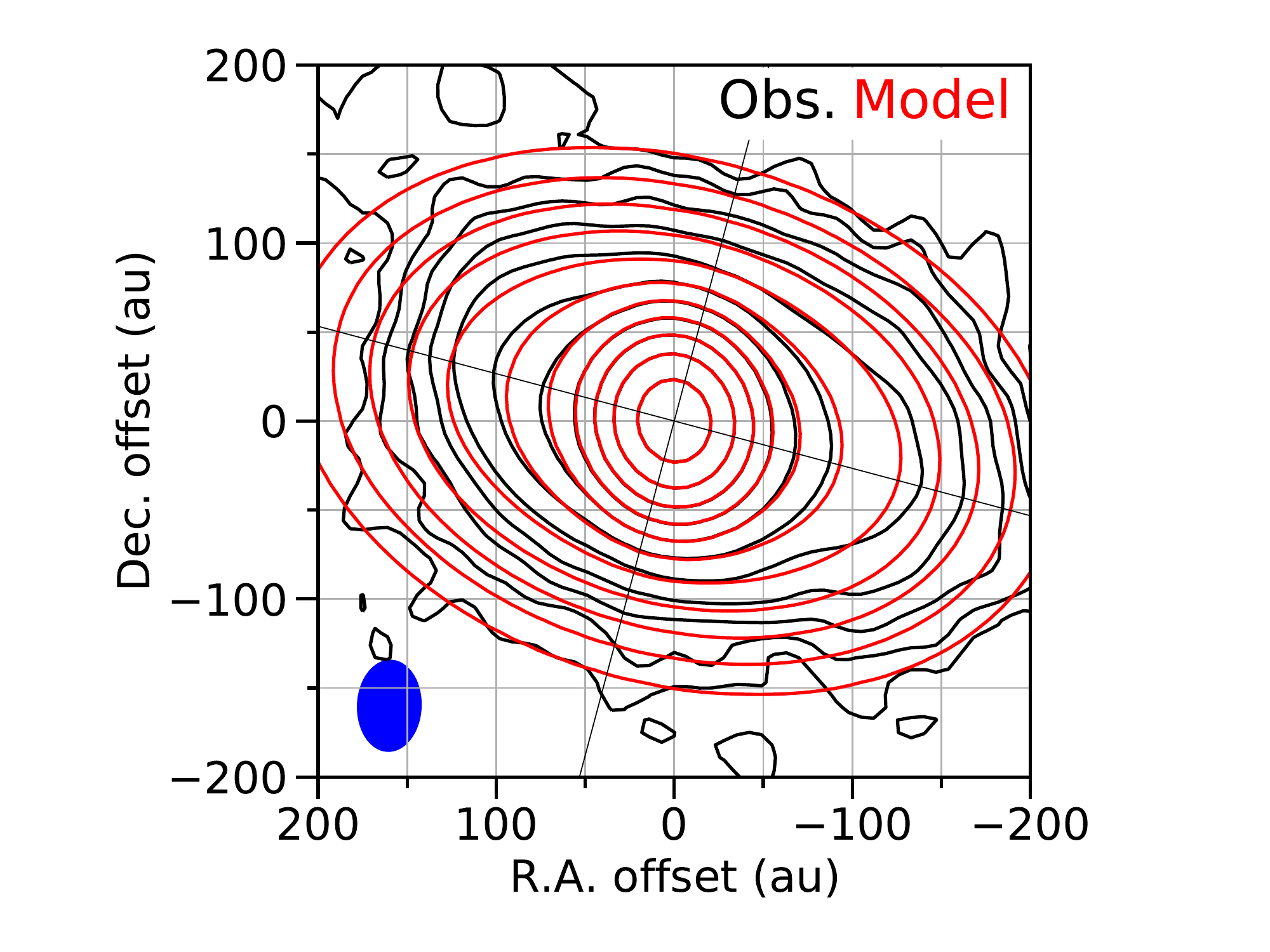}{0.42\textwidth}{(a)}
\fig{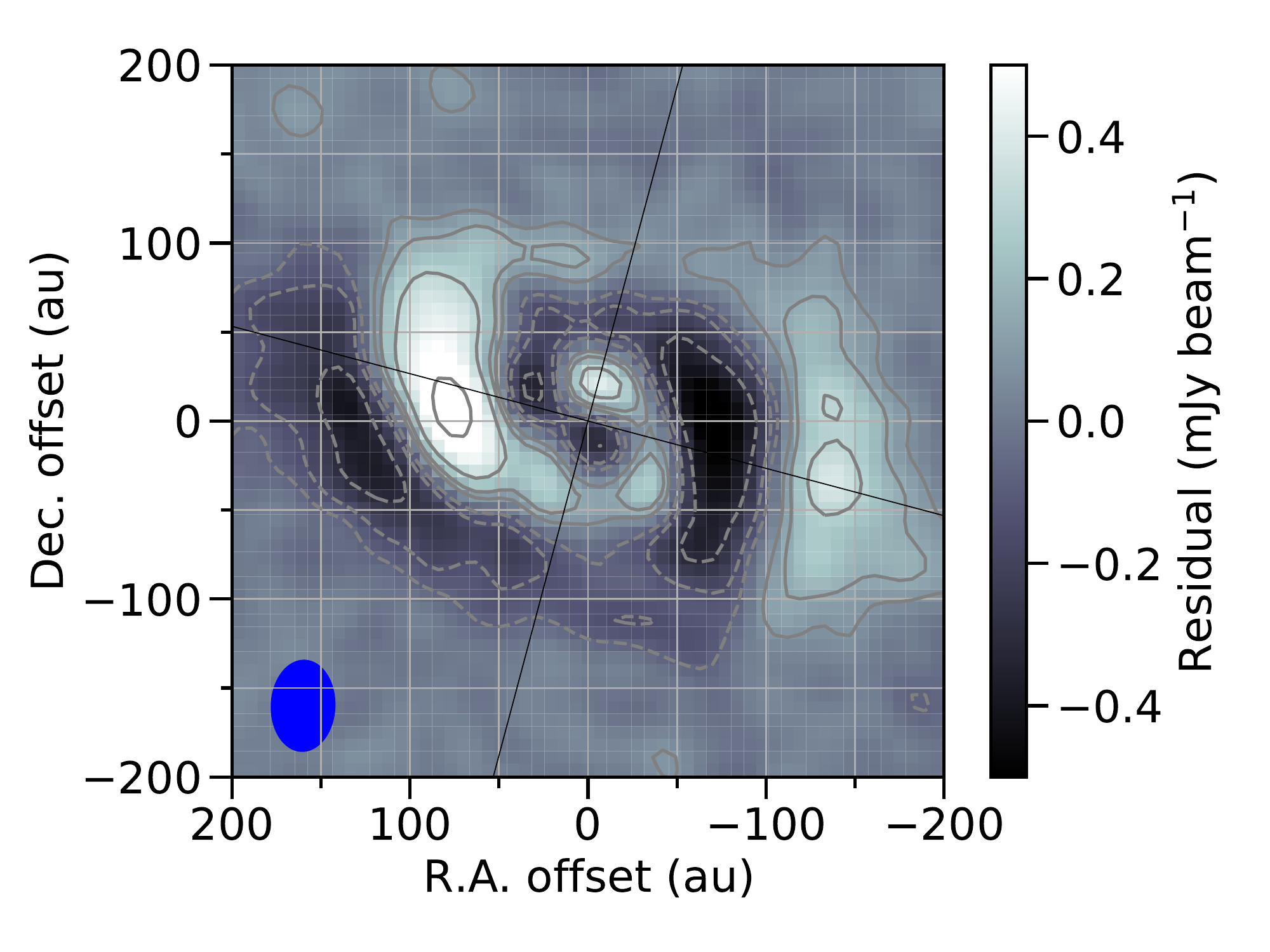}{0.42\textwidth}{(b)}
}
\caption{(a) Comparison of the ALMA 1.3 mm continuum and an axisymmetric model. (b) Residual, the observation minus the axisymmetric model, in the gray-scale and contour maps. The contour levels, diagonal lines, and the filled ellipse in both panels are the same as those in Figure \ref{fig:almacont}.
\label{fig:sym}}
\end{figure*}

Figure \ref{fig:sym} shows the best-fit model with the observation and the residual, the observation minus the best-fit model. Although the best-fit model reproduces the overall shape of the observation, the residual interestingly shows positive and negative spiral-like patterns. Note that this residual is a relative structure from the axisymmetric component; in other words, a single one-armed spiral plus an axisymmetric component can provide such positive and negative patterns. The orientation angle $\theta_0=75\arcdeg$ is close to the disk major axis reported in previous works \citep{hars14, aso15, bjer16}, and thus we define this orientation angle as the disk major axis and the orthogonal direction as the disk minor axis of TMC-1A in this paper.

\begin{figure}[htbp]
\epsscale{1}
\plotone{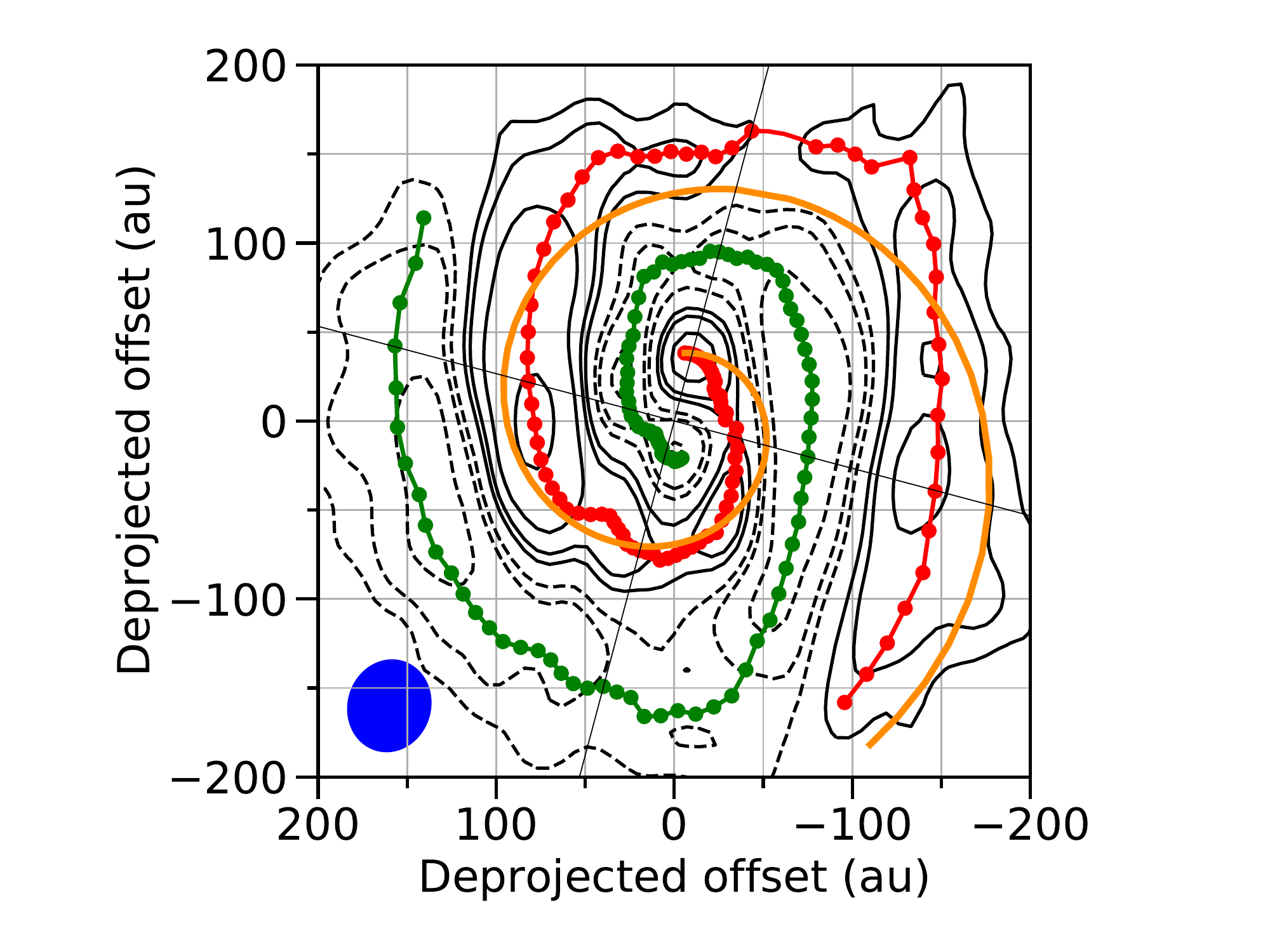}
\caption{The deprojected residual along the disk-minor axis, P.A.$=165\arcdeg$ by a factor of $\cos 53\arcdeg$. The filled ellipse is the deprojected beam, $0\farcs 38\times 0\farcs 34\ (-17.2\arcdeg)$. The diagonal lines are the same as those in Figure \ref{fig:almacont}. The red points are the intensity-mean radius at every $5\arcdeg$ of the positive residual, while the red line is an interpolated line. The green points and line are the contour parts for the negative residual. The orange curve is the logarithmic spiral fitted to the red points, $r=128\ {\rm au}\ \exp (-0.195\theta)$, where $r$ and $\theta$ are the radius and azimuthal angle in the deprojected plane, and $\theta$ is in the unit of radian.
\label{fig:dprj}}
\end{figure}

To investigate the spiral-like residual in more detail, we determine the relation between the radius and azimuthal angle along the positive and negative patterns. The intensity-weighted mean position (radius) is derived along the radial direction at every $5\arcdeg$. This mean radius uses intensities above $+3\sigma$ for the positive pattern and below $-3\sigma$ for the negative pattern. The derived mean radii are plotted in Figure \ref{fig:dprj} on the residual map de-projected using the orientation and inclination angles, $\theta_0=75\arcdeg$ and $i=53\arcdeg$. The red and green points are the mean radii for the positive and negative patterns, respectively. Figure \ref{fig:dprj} shows that the derived points represent the spiral-like patterns well. The mean radii for the positive pattern (red points) are fitted with a simple logarithmic spiral, $r=r_0~\exp (\alpha \theta)$, on the de-projected plane, where $r$ and $\theta$ are the polar coordinates, while $r_0$ and $\alpha$ are free parameters. The best-fit model is obtained by minimizing $\sum (r_{\rm obs}-r_{\rm mod})^2$ on the angle grid, where $r_{\rm obs}$ and $r_{\rm mod}$ are the observed mean radii (red points in Figure \ref{fig:dprj}) and the model fit, respectively.
The best-fit parameters are $r_0=128$~au and $\alpha=-0.195$. This logarithmic spiral is consistent with the positive pattern as shown in Figure \ref{fig:dprj} (see the orange curve). 

\begin{figure*}[htbp]
\epsscale{1.17}
\plotone{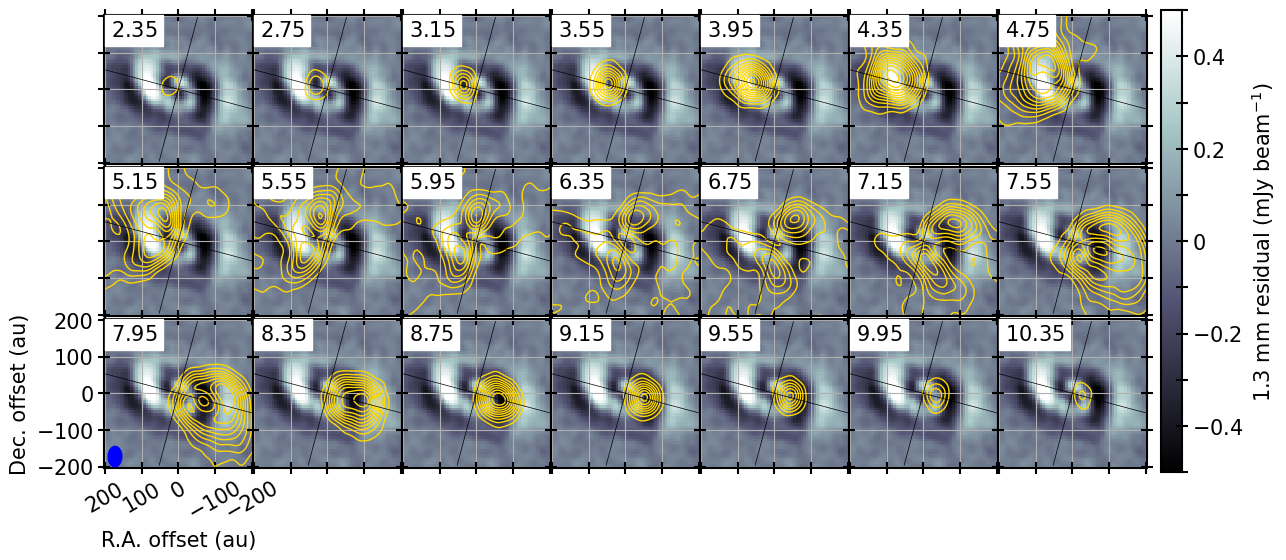}
\caption{Channel maps of the C$^{18}$O line with contours overlaid on the gray-scale map of the continuum residual. The contour levels are in $10\sigma$ steps, where $1\sigma$ corresponds to $1\ \mJB$. The diagonal lines are the same as those in Figure \ref{fig:almacont}. The filled ellipse is the C$^{18}$O synthesized beam. The velocity of each channel is denoted on the top-left corner of each panel.
\label{fig:resc18o}}
\end{figure*}

The continuum residual also shows a correlation with the C$^{18}$O emission. Figure \ref{fig:resc18o} shows the residual map together with the C$^{18}$O channel maps at a velocity resolution of $0.4~\kms$. One clear correlation can be seen at 4.35--$5.15~\kms$ and 7.15--$7.95~\kms$. From 2.35 to 4.35~$\kms$, the emission peak moves outward from the protostellar position along the disk major axis. The emission peak arrives at the eastern edge of the positive residual at $4.35~\kms$ and is divided into two peaks in the disk minor axis direction from $4.75~\kms$, not going beyond the eastern edge of the positive residual. Similarly, from 10.35 to $8.35~\kms$, the emission peak moves outward from the protostellar position along the disk major axis. The emission peak arrives at the western edge of the positive residual at $7.95~\kms$ and is divided into two peaks in the disk minor axis direction from $7.55~\kms$. These results on the eastern and western sides suggest that the C$^{18}$O emission is also concentrated inside the continuum residual, making the same asymmetry as the continuum residual. The high velocities $V<4.35~\kms$ and $V>8.35~\kms$ correspond to $|V-V_{\rm sys}|\gtrsim 2~\kms$. The fact that the emission peak is located along the disk major axis suggests that the C$^{18}$O emission traces the Keplerian disk around TMC-1A at this velocity range. Furthermore, the emission peak is divided at the lower velocities. This change at $|V-V_{\rm sys}|\sim 2~\kms$ suggests that the velocity channels at $|V-V_{\rm sys}|\sim 2~\kms$ trace the C$^{18}$O emission arising from the outer edge of the TMC-1A disk. In other words, the lowest rotational velocity in the TMC-1A disk (at the outer edge) is $\sim 2~\kms$. This is consistent with the disk radius, $\sim 100$~au, and the central protostellar mass, $\sim 0.7~\Msun$, of TMC-1A previously reported by \citet{aso15}.

\begin{figure}[htbp]
\epsscale{1}
\plotone{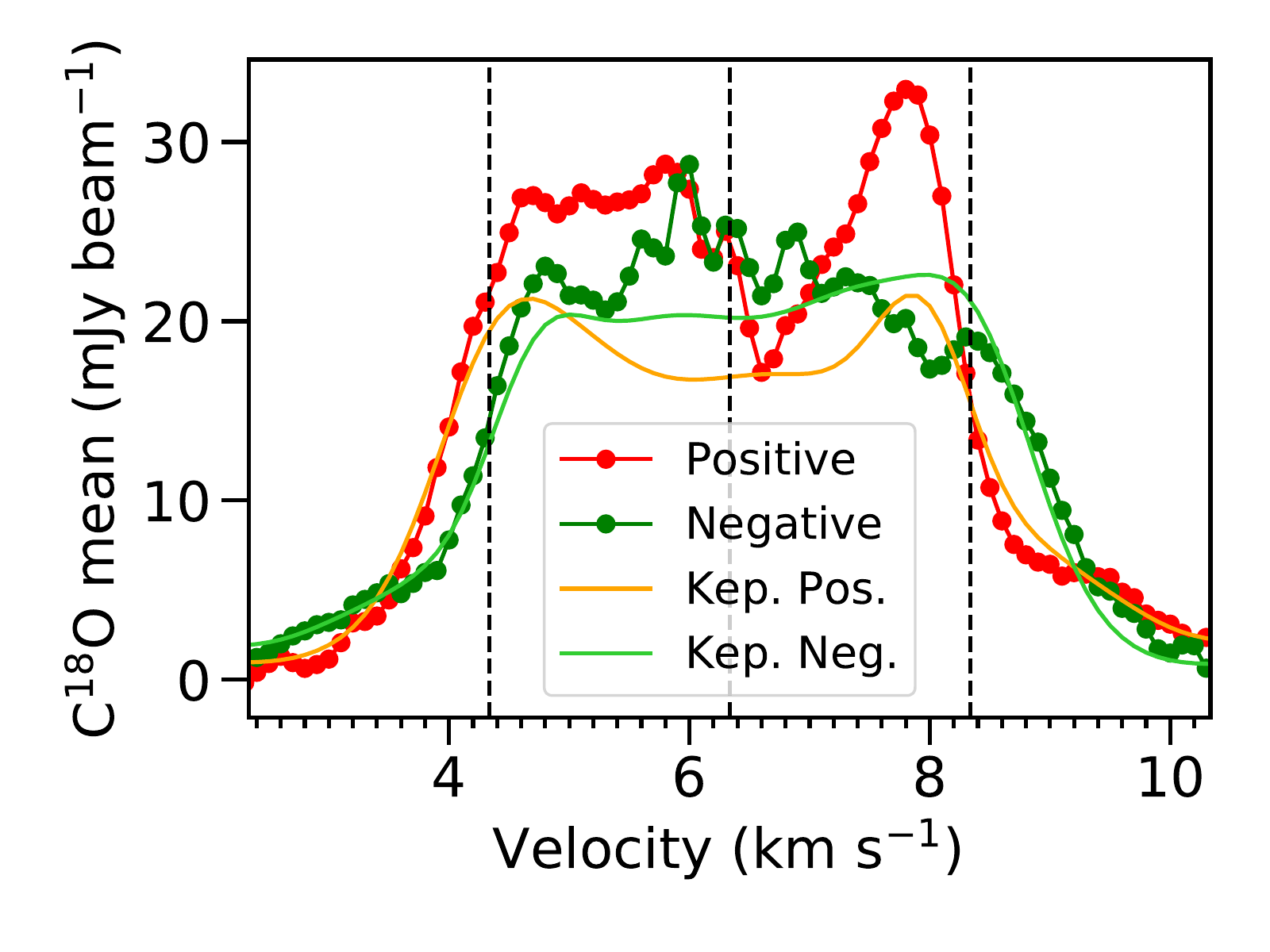}
    \caption{Mean intensities of the C$^{18}$O line along the positive and negative residuals of the continuum emission as each velocity channel. The vertical lines are $V_{\rm sys}-2\ \kms$, $V_{\rm sys}$, and $V_{\rm sys} + 2\ \kms$. The light color profiles show the mean intensities of the Keplerian disk model discussed in Section \ref{sec:kep}.
\label{fig:meanspi}}
\end{figure}

The C$^{18}$O intensity also implies another correlation with the continuum residual. Figure \ref{fig:meanspi} shows the mean intensity of the C$^{18}$O line on the positive (red in Figure \ref{fig:dprj}) and negative (green in Figure \ref{fig:dprj}) patterns at each velocity channel. The mean intensity on the positive pattern is significantly stronger than the negative one within $|V-V_{\rm sys}|<2~\kms$. This suggests intensity enhancement in the C$^{18}$O line at the same spiral as the intensity enhancement in the continuum emission, in this velocity range. On the other hand, the intensity on the positive pattern is stronger at blueshifted higher velocities ($V-V_{\rm sys}<-2~\kms$), while the intensity on the negative pattern is stronger at redshifted higher velocities ($V-V_{\rm sys}>2~\kms$). These differences at higher velocities are consistent with the Keplerian rotation as discussed in more detail in Section \ref{sec:kep} below using a Keplerian disk model; the solid lines without points in Figure \ref{fig:meanspi} are from the model in Section \ref{sec:kep}.

\subsection{Keplerian Rotation and Infall Motion} \label{sec:kep}

The analysis in the previous subsection revealed correlations between the spatial distributions of the C$^{18}$O emission and the continuum residual. In this subsection, the velocity structure of the C$^{18}$O emission is modeled to help interpret the correlations. 
TMC-1A is known to have a Keplerian disk \citep{hars14, aso15, bjer16}. Hence, we model the Keplerian disk by fitting with a toy model. The main purpose of this fitting is not to constrain physical conditions of the disk but to derive the distribution of intensity arising from the disk in the position-position-velocity space. The velocity field in the toy model is the Keplerian rotation determined by the central protostellar mass $M_*$ and has a uniform line profile described as a Gaussian function with a standard deviation of $c_s$. The systemic velocity is $V_{\rm sys}$. The intensity in the model disk is a power-law function of radius up to an outer radius $R_{\rm out}$, $I_{100} (r/100~{\rm au})^{-p}$, where $r$, $I_{100}$, and $p$ are the radius on the disk plane, a coefficient, and a power-law index, respectively. This intensity field is located on two parallel planes at $\pm H$ from the midplane, to mimic the scale height of the disk. The model disk is oriented by $\theta _0$ and inclined by $i$: the angles derived from the fitting to the continuum emission (Section \ref{sec:sym}). This model intensity is convolved with the Gaussian beam having the same FWHM as the C$^{18}$O observation. In summary, the toy model has seven free parameters, $M_*$, $c_s$, $V_{\rm sys}$, $I_{100}$, $p$, $R_{\rm out}$, and $H$. The best-fit parameters are derived by minimizing $\chi^2 = \sum (f_{\rm obs}-f_{\rm mod})^2/\sigma ^2$ over the velocity range of 1.4--4.4 and 8.4--$11.4~\kms$ in the image domain, where $f_{\rm obs}$. $f_{\rm mod}$, and $\sigma$ are the observed C$^{18}$O intensity, model intensity, and rms noise level of the C$^{18}$O observation. The best-fit model has the parameters of $(M_*,\ c_s,\ V_{\rm sys},\ I_{100},\ p,\ R_{\rm out},\ H)=(0.72~\Msun,\ 0.38~\kms,\ 6.34~\kms,\ 1.5~{\rm mJy~pixel}^{-1},$ $0.48,\ 150~{\rm au},\ 14~{\rm au})$. The central protostellar mass is consistent with the previous work \citep[$0.68~\Msun$;][]{aso15}. The best-fit model well reproduces the observed C$^{18}$O channel maps in the fitted velocity range, as shown in Appendix \ref{sec:chmod}. This model also reproduces the mean intensities along the positive and negative residuals in the fitted velocity range, $|V-V_{\rm sys}|>2.0~\kms$, as shown in Figure \ref{fig:meanspi}.

\begin{figure*}[htbp]
\gridline{
\fig{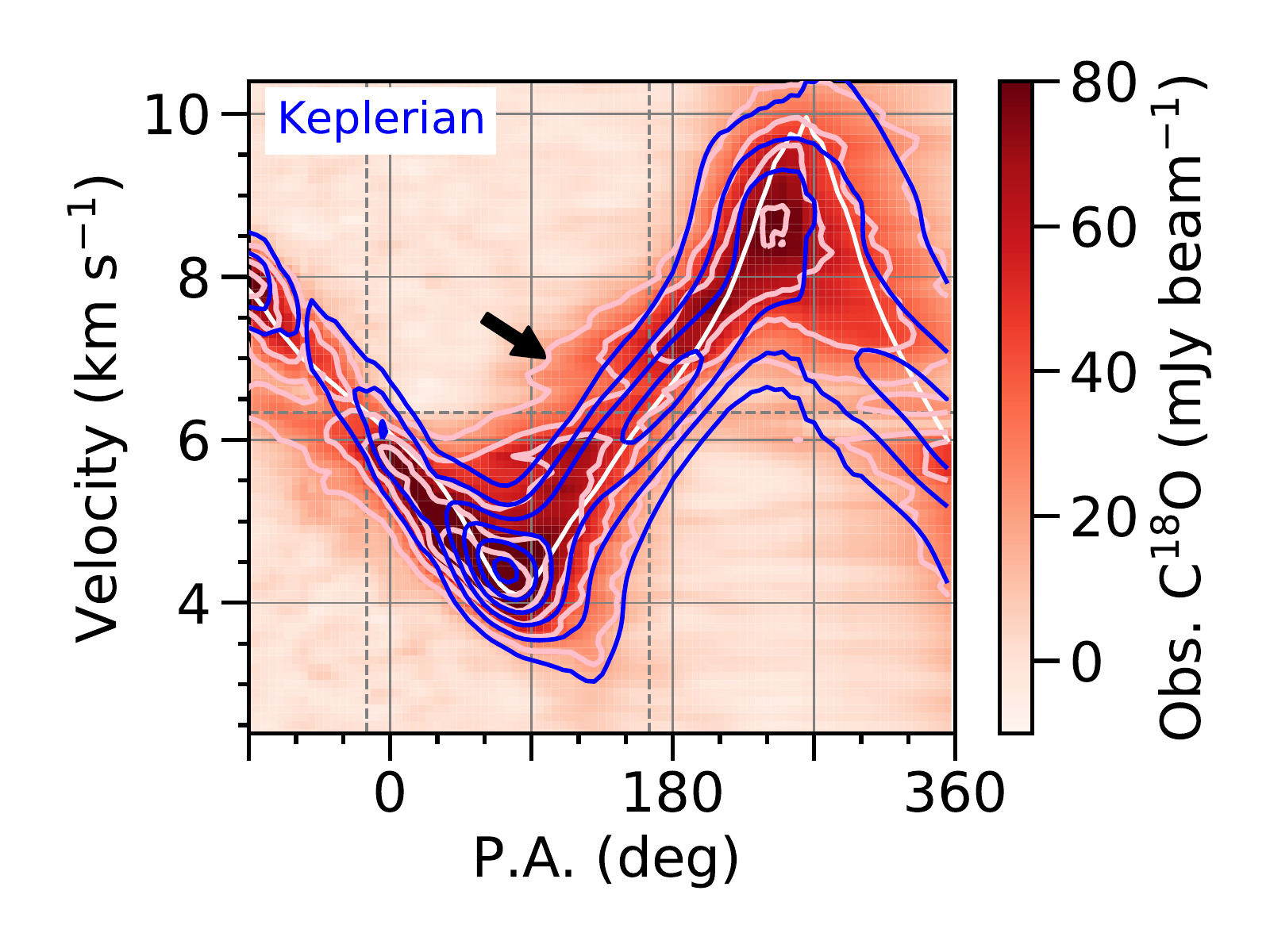}{0.4\textwidth}{(a)}
\fig{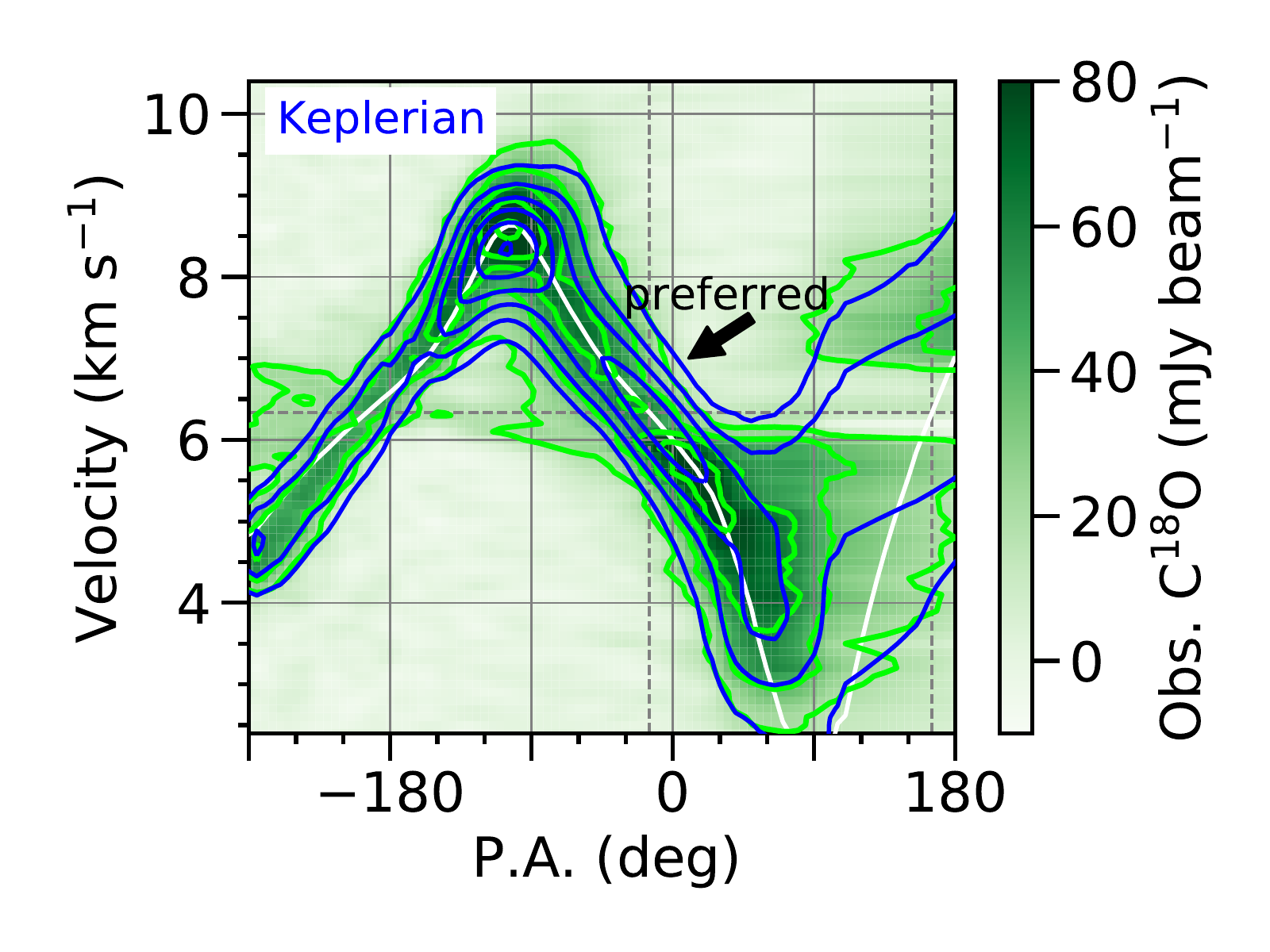}{0.4\textwidth}{(b)}
}
\gridline{
\fig{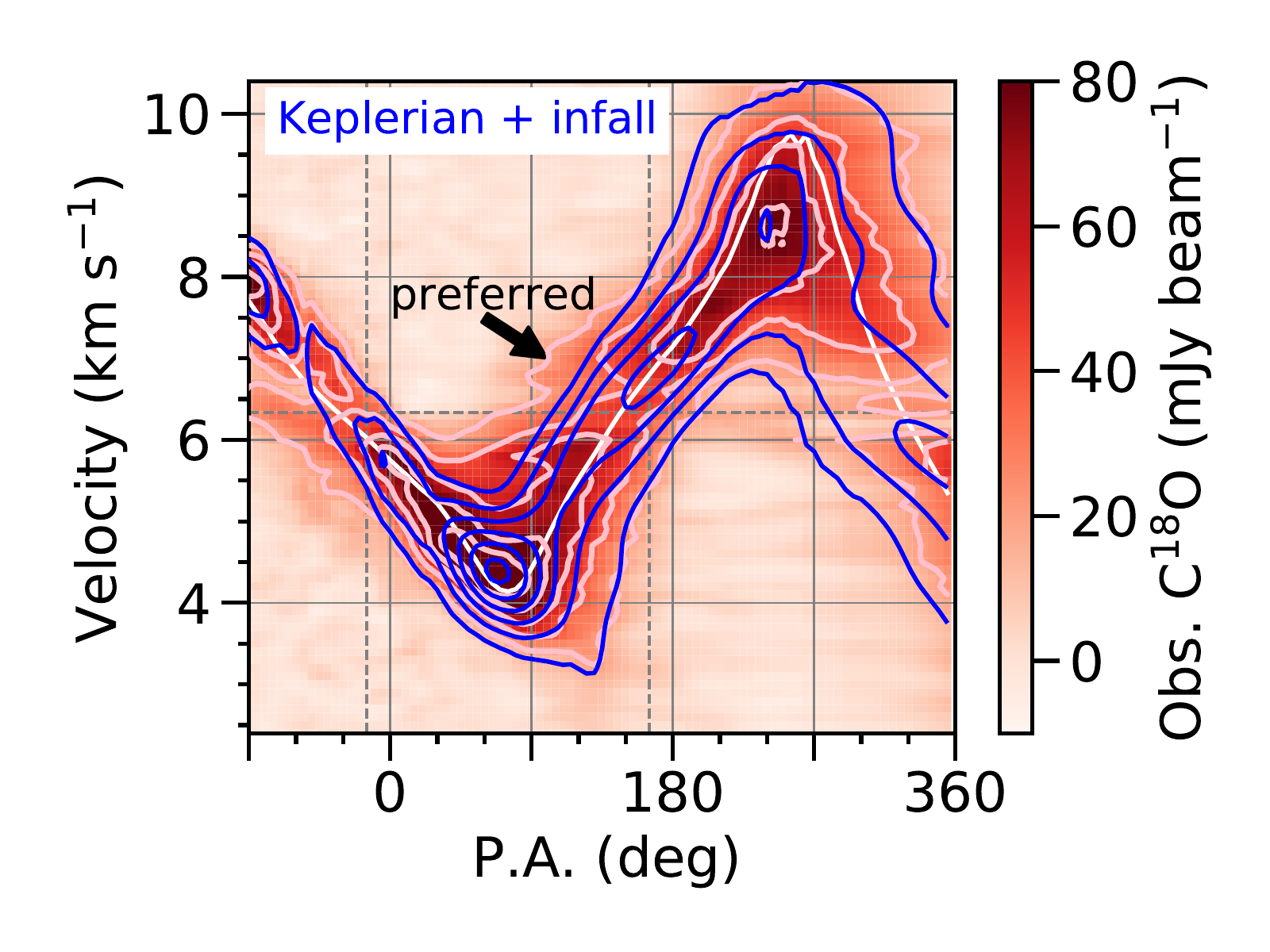}{0.4\textwidth}{(c)}
\fig{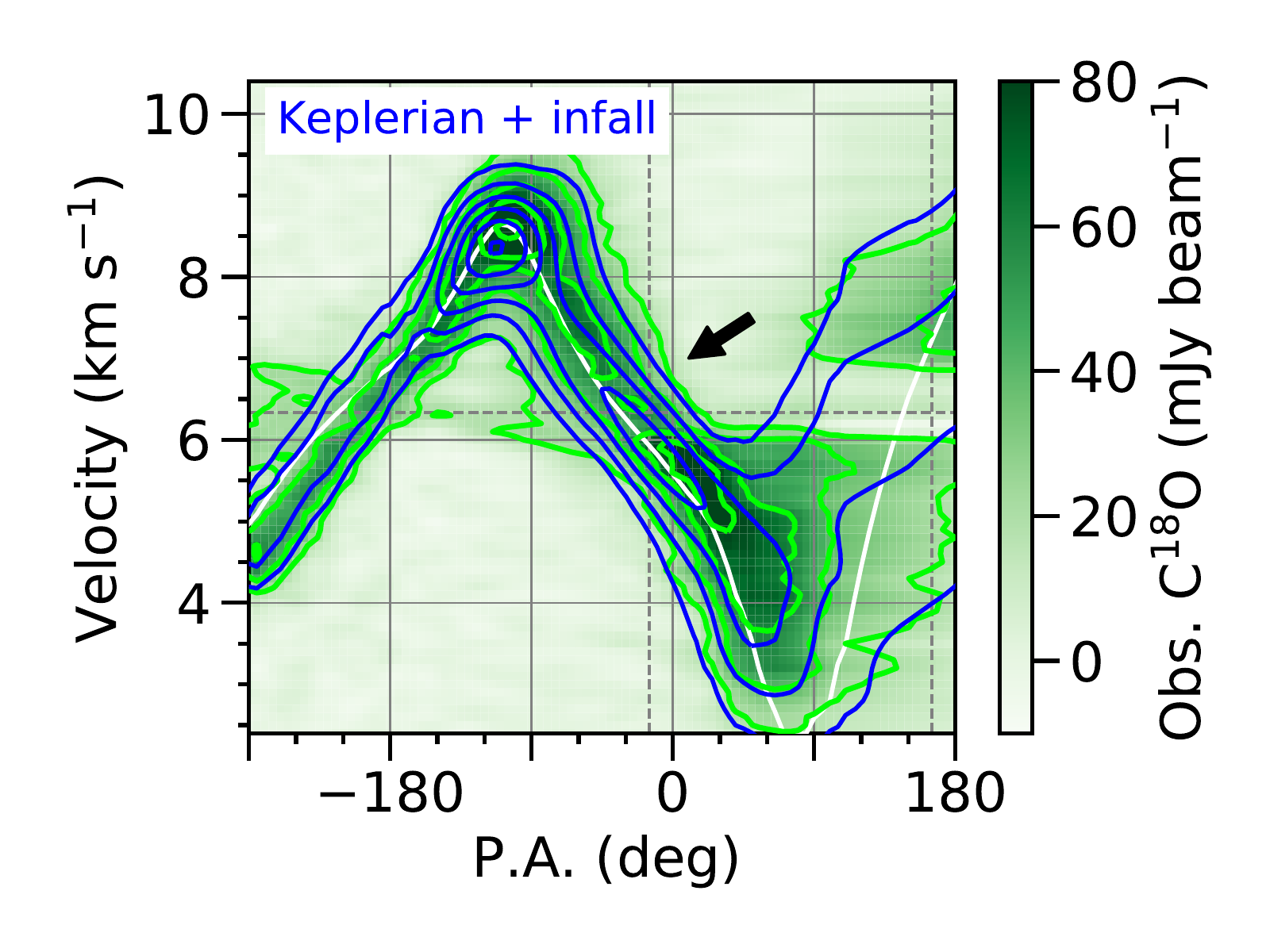}{0.4\textwidth}{(d)}
}
\caption{Position-velocity diagrams along the positive (left) and negative (right) residuals of the continuum emission. The abscissa value is the position angle from the north to the east. The blue contours show a Keplerian disk model or a Keplerian$+$infall model. The contour levels are in $10\sigma$ steps. The white curve is the velocity field on the midplane. The vertical and horizontal dashed lines denote the disk-minor axis and the systemic velocity, respectively. The arrows point to the main difference between the two models. Note that the positive residual prefers the Keplerian$+$infall model, while the negative residual is better fitted by the Keplerian-only model.
\label{fig:pvsp}}
\end{figure*}

The Keplerian disk model can be used to inspect non-Keplerian components in the C$^{18}$O data cube. In particular, it is worth comparing the velocity structure along the positive and negative residuals in the continuum emission with the Keplerian velocity field. For this purpose, Figure \ref{fig:pvsp}(a) and \ref{fig:pvsp}b show the position-velocity diagrams of the observed C$^{18}$O emission and the Keplerian disk model along the positive (red in Figure \ref{fig:dprj}) and negative (green in Figure \ref{fig:dprj}) residuals, where the abscissa coordinate is the position angle measured from the north to the east. These diagrams show that the overall velocity structure in the C$^{18}$O line is the Keplerian rotation. A possible offset of the emission peak can be found in the positive-residual diagram around the minor axis (P.A.$=165\arcdeg$) as highlighted with an arrow in Figure \ref{fig:pvsp}(a): the observed velocity structure appears slightly redshifted with respect to the Keplerian disk model by 0.5--$0.6~\kms$. This angle corresponds to a radius of 70--80~au on the de-projected plane (Figure \ref{fig:dprj}). The Keplerian velocity at this radius is 2.8--$3.0~\kms$ with $M_*=0.72~\kms$. The ratio between the offset $\sim 0.5~\kms$ and the Keplerian velocity 2.8--$3.0~\kms$ is consistent with the logarithmic spiral with $\alpha \sim 0.2$ derived in Section \ref{sec:sym}, suggesting a flowing motion along the positive residual with a radial infall velocity of $\sim 0.2$ times the rotational velocity. The presence of the infalling component is consistent with the velocity gradient in the minor-axis direction mentioned in Section \ref{sec:line}.

Figure \ref{fig:pvsp}c shows comparison between the observed C$^{18}$O emission (same as in Figure \ref{fig:pvsp}a) and an infalling rotating model that has the Keplerian rotation and a radial infall velocity of 0.2 times the Keplerian rotation. The emission peak of this Keplerian$+$infall model traces the observed peak better around the minor axis than that of the Keplerian disk model, while other parts are similar to each other. In contrast, the emission peak of the Keplerian disk model is more consistent with the observed peak in the negative-residual diagram than that of the Keplerian$+$infall model (Figure \ref{fig:pvsp}b and \ref{fig:pvsp}d), in particular, around the minor axis (P.A.$=-15\arcdeg$). This difference implies that only the positive residual has an inflow motion, supporting the possibility that a single one-armed spiral provides both positive and negative patterns as a relative structure with respect to the axisymmetric average component as mentioned in Section \ref{sec:sym}.

The radial infall velocity at $r=70$-80~au revealed in this subsection corresponds to $\sim 14\%$ of the free fall velocity, since the free fall velocity is $\sqrt{2}$ times the Keplerian velocity. Although the radial infall velocity at outer radii cannot be evaluated from our results, a larger ratio of $\sim 30\%$ is reported in \citet{aso15} at $r=100$-200 au. This difference may suggest soft landing of the collapsing envelope material, through the one-armed spiral, onto the outer part of the disk.

\clearpage

\section{Discussion} \label{sec:discussion}
The origin of polarization is not as simple on the disk scale around protostars as on the envelope scale and larger. Polarization on the larger scales, such as that probed by our SMA observations, is generally thought to come from magnetic alignment. We discuss possible origins of the ALMA polarization, as well as that of the accretion flow, in this section.

\subsection{Central Polarization} \label{sec:cpol}      
\begin{figure}[htbp]
\epsscale{1}
\plotone{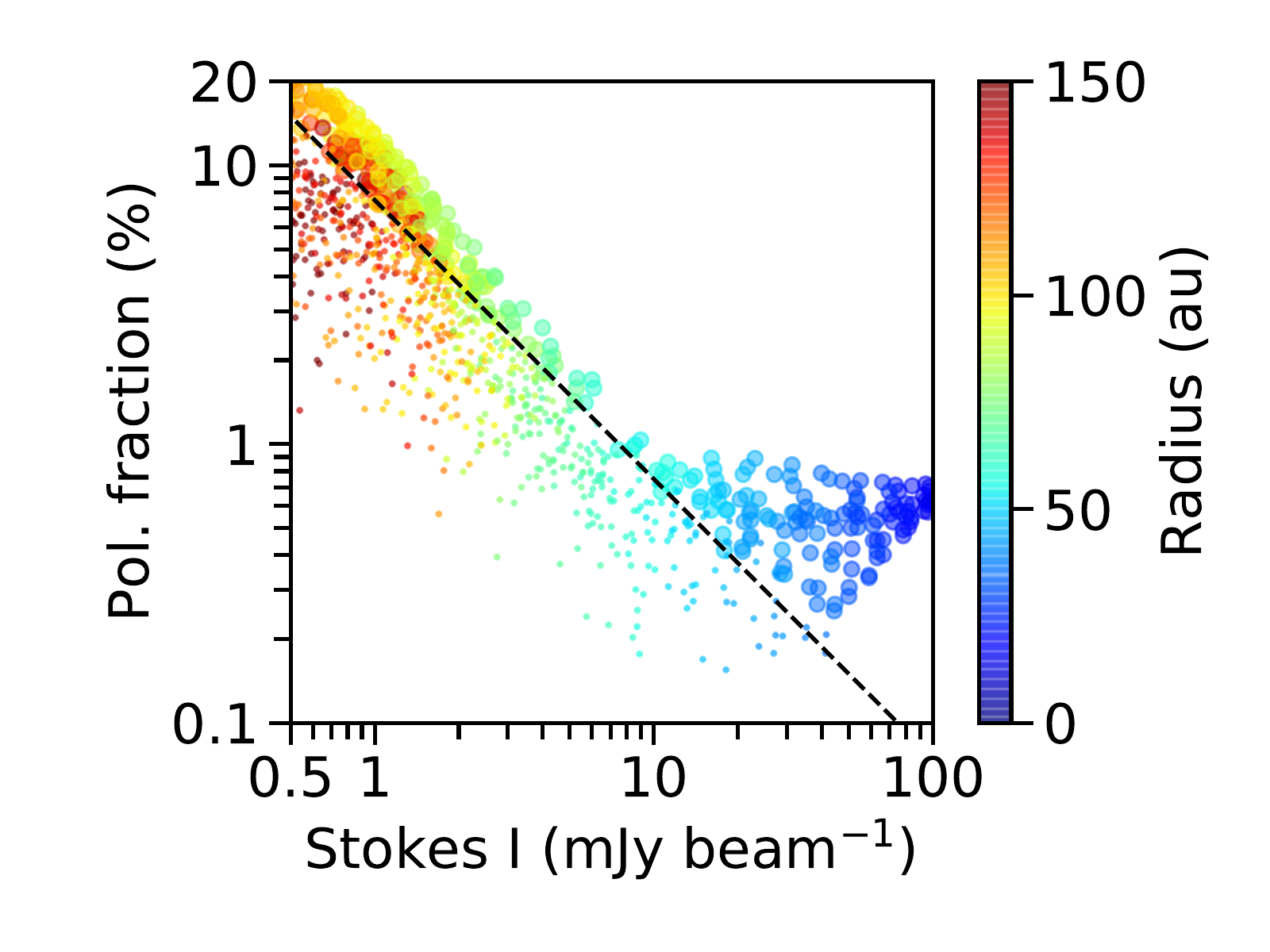}
\caption{Polarization fraction as a function of Stokes I of the ALMA 1.3 mm continuum emission. The color denotes the radius of each point. The dashed line is the $3\sigma$ detection level for the polarization intensity. The data points with the larger point size correspond to the positions where de-biased polarization angles are derived in Figure \ref{fig:almacont}.
\label{fig:polfi}}
\end{figure}

The ALMA polarization shows two components clearly divided by a de-polarized ring: central ($r\lesssim 50$~au) and a northern/southern component (Figure \ref{fig:almacont}). The central component shows polarization E-vectors along the disk minor-axis near the axis and include an azimuthal component near the outer edge. These two morphological features suggest dust self-scattering occurring on the disk surface as the origin of the central polarization. Polarization due to self-scattering is detected in other Class 0 and I protostars \citep[e.g., ][]{lee18}. Theoretical simulations of self-scattering predict such polarization directions and polarization fractions $\sim 0.8\%$ to $\sim 1.0\%$ with inclination angles of $50\arcdeg$ to $60\arcdeg $ \citep{yang17}, which is also consistent with the observed polarization fraction. Figure \ref{fig:polfi} shows the observed polarization fraction as a function of Stokes $I$. At the radii of the central component $r\lesssim 50$~au, the polarization fraction is independent from Stokes $I$. This tendency prefers self-scattering to the magnetic alignment. The theoretical simulations suggest that self-scattering requires a high optical depth ($\tau \gtrsim 1$) to produce a polarization fraction high enough ($\gtrsim 0.5\%$) to be observed. Previous observations toward TMC-1A indeed reported that the 1.3 mm dust continuum emission is optically thick in a central region with $r\lesssim 20$--30~au \citep{hars18}.

The high optical depth in the TMC-1A disk was interpreted as a result of a high opacity due to dust grains with mm size \citep{hars18}. Although a massive disk can also produce a high optical depth, such a disk should be gravitationally unstable and thus show some sign of instability on the disk (e.g., spiral arm or fragmentation). Previous observational studies did not show such a sign in the TMC-1A disk at the mm wavelength, preferring the mm grain scenario to the massive disk scenario. Millimeter grains are, however, not consistent with the self-scattering origin of polarization because self-scattering requires a maximum grain size of $\sim 80$--$300~\micron$ to produce a polarization fraction of $\gtrsim 1\%$, while mm-sized grains decrease the predicted polarization fraction to $\sim 0.01\%$ \citep{kata16}. In addition to the self-scattering-induced polarization, we have revealed a spiral-like residual in the 1.3 mm continuum emission. This structure may hint at the gravitational instability; hence, the high optical depth in the TMC-1A disk could be due to a massive disk. For these reasons, we suggest that dust grains mainly have a size up to a few 100s~$\micron$ in the TMC-1A disk.

\subsection{Northern and Southern Polarization} \label{sec:nspol}
The northern and southern components of the ALMA polarization are unlikely produced by scattering, because of their high polarization fraction, location in the outer part of the disk along the minor axis, and relatively low optical depth. We consider three alternative possibilities. The first possibility is magnetic alignment by toroidal magnetic fields in the disk or outflow associated with TMC-1A. A toroidal field is described by an ellipse elongated along the disk major-axis, which is broadly consistent with the $90\arcdeg$-rotated vectors detected with ALMA except for the central component. The polarization fraction in the north/south component is $\sim 10\%$ and shows an anticorrelation with Stokes $I$ (Figure \ref{fig:polfi}). Such a high polarization fraction is possible with elongated dust particles aligned in a magnetic field, although boundary areas of a structure in interferometric observations can show a relatively high polarization fraction due to a larger filtering effect in Stokes $I$ \citep{kwon19}. \citet{kwon19} also reported a negative power-law index, $-0.4$ to $-1.0$, between the two quantities on a few 1000 au scale in the protostellar system L1448 IRS 2, where polarization originates in the magnetic alignment. Because the TMC-1A system is inclined from the line-of-sight, the polarization near the minor axis is expected to be stronger than near the major axis because non-spherical grains rapidly spinning along their magnetically aligned short axis are effectively oblate spheres and such aligned oblate spheres are seen more edge-on near the minor axis and more face-on (and thus more circular with lower polarization) near the major axis (see the top panels of Figure 10 of \citet{ch.la07} or the middle panels of Figure 3 of \citet{yang16}). In addition, because of the disk inclination, the toroidal fields show a low (high) curvature near (far from) the outflow axis on the projected plane. When magnetic fields have a high curvature, observed polarization can have a low polarization fraction because signals with different polarization directions are summed in an observational beam and cancelled out. These two effects could explain the fact that the polarized emission is detected with ALMA mainly in the north and south of the central protostar.

The second possibility is the mechanical alignment called ``Gold mechanism" by the associated outflow \citep{gold52}. Dust grains in the outflow lobe are aligned in the outflow direction by the gas motion, as if each grain is a boat in a river. The aligned grains, then, produce the polarization direction parallel to the outflow direction as observed in our ALMA observations. The TMC-1A outflow shows a projected velocity faster than $\sim 10~\kms$, indicating that the velocity in this outflow is sufficiently high to provide supersonic gas flow that makes the Gold mechanism efficient \citep{laza94}. However, because the disk material is expected to be much denser than the outflow material, most of dust emission must be coming from the disk. If the disk emission is not polarized, the intrinsic polarization from the outflow emission must be much higher than the observed 10\%, which is unlikely. Furthermore, while the blue-shifted outflow spatially coincides with the northern polarization component, there is no clearly detected red-shifted outflow that spatially coincides with the southern polarization component. For these reasons, we regard the Gold mechanism as less likely. We should mention that another version of mechanical alignment was studied by \citet{hoan18}, where the grains are expected to align their long axes perpendicular to the gas flow. It would predict polarization E-vectors perpendicular to the minor axis, which are not observed in this component.

\begin{figure}[htbp]
\epsscale{1}
\plotone{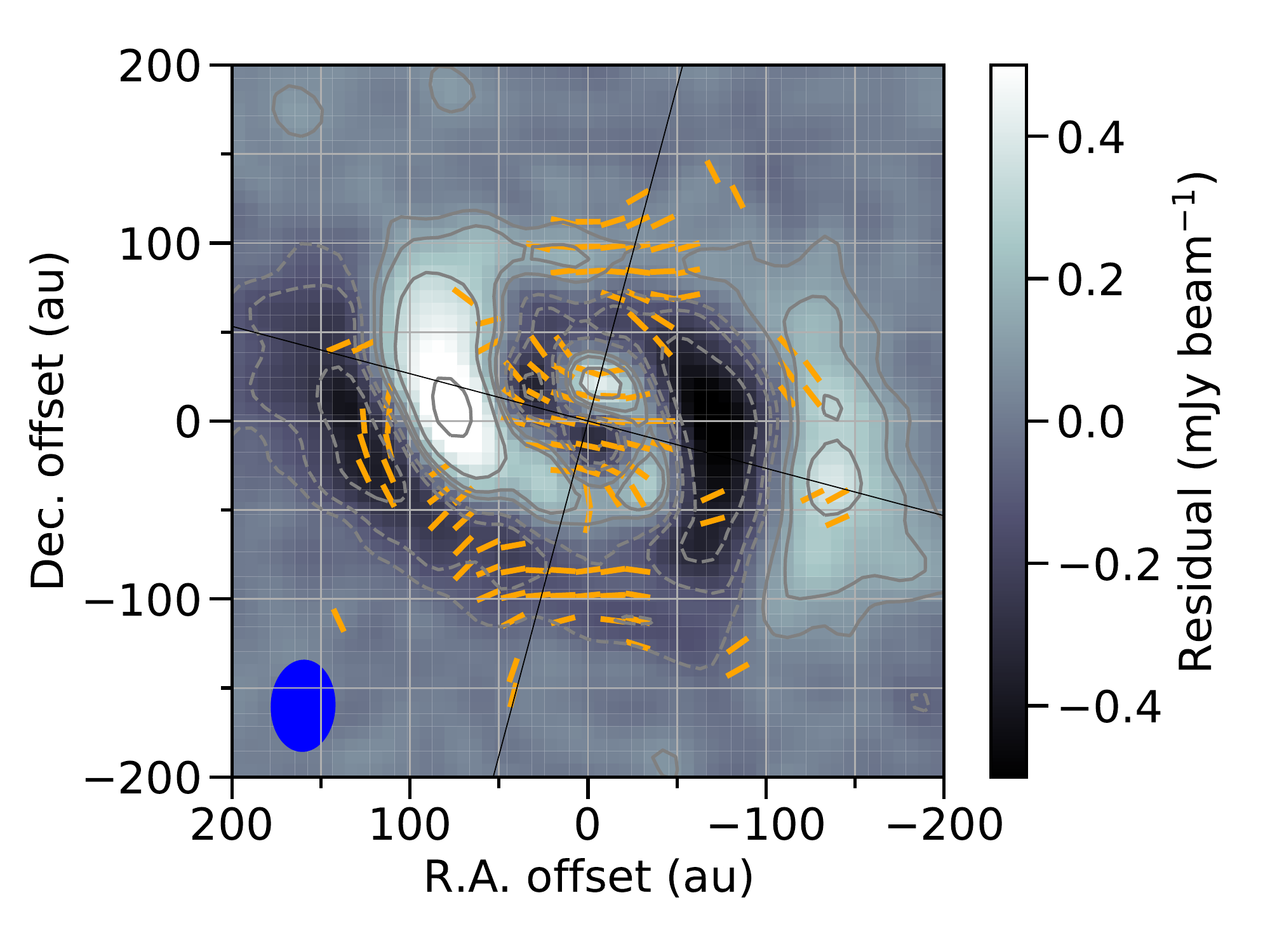}
\caption{ALMA polarization directions rotated by $90\arcdeg$ overlaid on the spiral-like residual in the 1.3 mm Stokes $I$. The northern polarization shows the vectors ($90\arcdeg$-rotated) along the spiral-like residual.
\label{fig:res90}}
\end{figure}

The third possibility is magnetic alignment by a magnetic field along the accretion flow suggested in Section \ref{sec:kep}. This possibility is motivated by the fact that the northern polarization is overlapped on the spiral-like residual and the $90\arcdeg$-rotated vectors are also along the residual, as shown in Figure \ref{fig:res90}. This peculiar structure could explain the limited location of the detected polarization. This possibility cannot be applied to the southern polarization because the southern polarization is overlapped on the negative spiral outside the one-armed spiral.

\subsection{DCF Method to the ALMA Polarization} \label{sec:dcfalma}
\begin{figure*}[htbp]
\gridline{
\fig{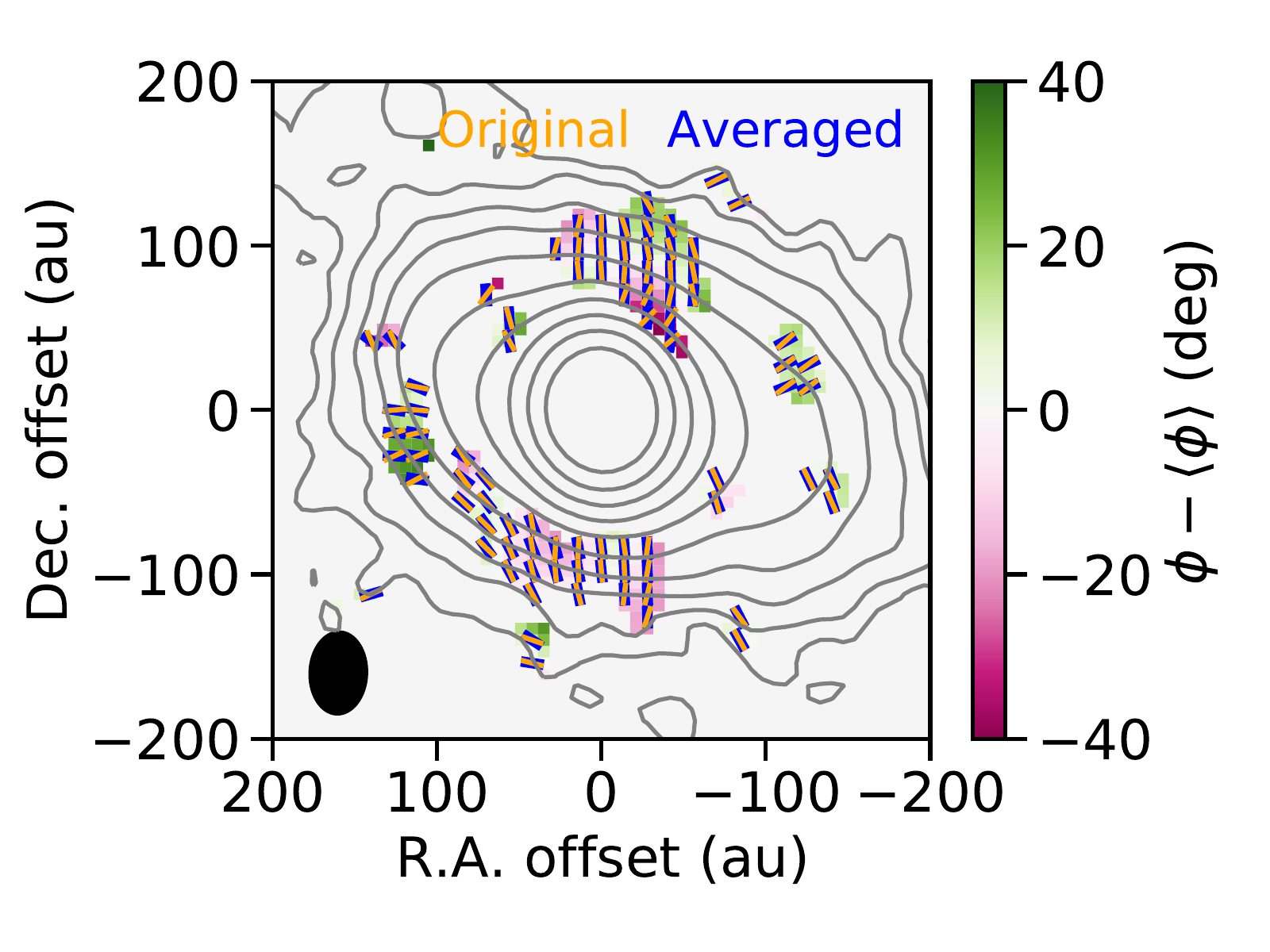}{0.42\textwidth}{(a)}
\fig{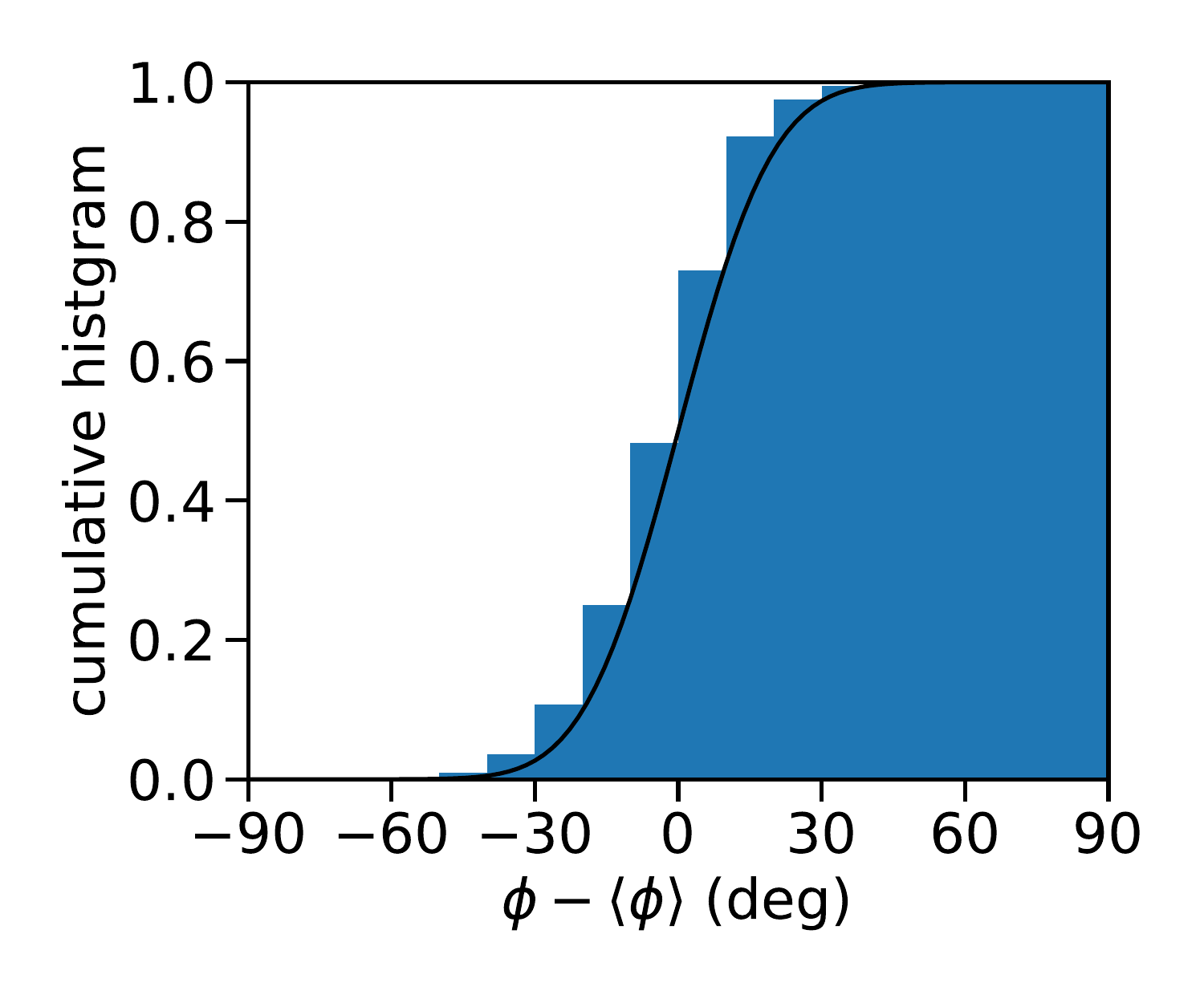}{0.37\textwidth}{(b)}
}
\caption{(a) Deviation of polarization angles observed with ALMA from 2-beam averaged angles. The blue segments denote the 2-beam averaged angles $\langle \phi \rangle$. The orange segments, the contour map, and the filled ellipse are the same as those in Figure \ref{fig:almacont}. The coordinates are relative to the protostellar position. (b) Cumulative histogram of the relative polarization angle from the 2-beam averaged angle, $\phi - \langle \phi \rangle$. The black curve is the error function with a standard deviation of $16\arcdeg$.
\label{fig:dcfalma}}
\end{figure*}

The northern and southern components in the ALMA polarized 1.3 mm emission are likely to originate in toroidal magnetic fields in TMC-1A, as discussed in the previous subsection. Then, the DCF method allows us to roughly estimate the field strength,
as it does for our SMA data (Section \ref{sec:dcf}). 
The dispersion of the polarization angle is measured from the 2-beam averaged angles using a 2D Gaussian function with a FWHM of $\sim 0\farcs 6$. Note that the central polarization is masked by a threshold of polarization fraction $<1\%$ and not used for the average-angle calculation. Figure \ref{fig:dcfalma}(a) shows the comparison of the original and average angles. The dispersion is calculated to be $\delta \phi=16\arcdeg$ in this case. Figure \ref{fig:dcfalma}(b) shows the cumulative histogram of the relative angle with the error function with a standard deviation of $16\arcdeg$. The northern and southern components have Stokes $I$ $\sim 0.6~\mJB$. This corresponds to a mass column density of 0.12--$0.33~{\rm g}~{\rm cm}^{-2}$ on the same dust opacity, temperature, and gas-to-dust mass ratio as in Section \ref{sec:dcf}. If this column density is distributed over the scale height of 14~au (Section \ref{sec:kep}), the mean density is $\langle \rho \rangle =0.6$--$1.6\times 10^{-15}~{\rm g}~{\rm cm}^{-3}$. The velocity dispersion is also in the same range 0.3--$1.0~\kms$ as in Section \ref{sec:dcf}. Consequently, from Equation (\ref{eq:dcf}), the field strength is calculated to be 5--25~mG.

\subsection{Accretion Flow}
The spiral-like residual suggests a one-armed accretion flow in the TMC-1A disk.
A simple interpretation of this flow is occasional mass accretion from the associated envelope. The C$^{18}$O emission also supports flowing motion along the spiral.
Similar morphology is also identified in multiple protostellar systems. \citet{tobi16} show a one-armed spiral with a radius of $\sim 200$~au in the triple protostellar system L1448 IRS3B observed in the 1.3 mm continuum emission. Their line observations in C$^{18}$O $J=2-1$ indicate a rotational motion perpendicular to an associated outflow. \citet{taka14} show a one-armed spiral with a radius of $\sim 200$~au in the protostellar binary L1551 NE observed in the 0.9 mm continuum emission. The authors reproduced the spiral structure and kinematics observed in the C$^{18}$O $J=3-2$ line using a hydrodynamic simulation. The similarity of the TMC-1A spiral with these multiple systems may hint at gravitational instability in the TMC-1A disk, although the instability may be weaker here so that the spiral can only be seen as the residual from the axisymmetric component.

An inner part of the spiral-like residual also appears to delineate a part of a ring with a radius of $\sim 50$~au. Such a ring may result from the `growth front' \citep[or the pebble production line;][]{la.jo14}, where dust grains drastically grow from $\micron$ size to mm size. The growth front is estimated to be 50-60~au in radius with the typical age of Class I protostars, 0.1~Myr \citep{ohas20}, which is consistent with the inner part of the spiral-like residual.

\section{Conclusions} \label{sec:conclusions}
We have observed the linearly polarized dust continuum emission at 1.3 mm in the Class I protostellar system TMC-1A using the SMA and ALMA at angular resolutions of $\sim 3\arcsec$ (400 au) and $\sim 0\farcs 3$ (40 au), respectively. The ALMA observations also included the CO, $^{13}$CO, and C$^{18}$O $J=2-1$ lines. The main results are summarized below.
\begin{enumerate}
    \item The SMA polarization observations trace magnetic fields in the TMC-1A envelope on a 1000-au scale. The field directions are between the parallel and perpendicular directions to the outflow axis. We estimate the field strength to be 1-5 mG by applying the DCF method to the SMA polarization. The diagonal direction and mG-order strength of the magnetic field are consistent with the previous prediction by \citet{aso15} to explain the radial infall velocity at $\sim 30\%$ of the free fall velocity.
    \item We subtract an axisymmetric component from the ALMA continuum emission, Stokes $I$, and discover a spiral-like residual for the first time in TMC-1A. The deprojected spiral suggests an accretion flow with an radial infall velocity at $20\%$ of the rotational velocity along the spiral-like residual. Comparison of the C$^{18}$O emission and a Keplerian disk model also supports this ratio between the rotational and radial infall velocities.
    \item The polarized emission observed with ALMA consist of a central component and a north/south component. The central component can be interpreted as a result of self-scattering because the polarization directions are mostly along the disk minor-axis, but with an azimuthal component near the major axis, and the polarization fraction is $\sim 0.8$\% independent of the polarization intensity.
    \item The north/south polarization component is located along the outflow direction (i.e., in the north and south of the protostellar position), and the polarization E-vectors are also broadly parallel to the outflow direction. Three potential mechanisms are discussed for this polarization: (1) toroidal magnetic fields in the outflow or disk in this system, (2) mechanical grain alignment by the gaseous outflow, and (3) a magnetically channeled accretion flow as suggested by the spiral-like residual in Stokes $I$.
\end{enumerate}

\acknowledgments
This paper makes use of the following ALMA data: ADS/JAO.ALMA\#2018.1.00701.S. ALMA is a partnership of ESO (representing its member states), NSF (USA) and NINS (Japan), together with NRC (Canada), MOST and ASIAA (Taiwan), and KASI (Republic of Korea), in cooperation with the Republic of Chile. The Joint ALMA Observatory is operated by ESO, AUI/NRAO and NAOJ.
The Submillimeter Array is a joint project between the Smithsonian Astrophysical Observatory and the Academia Sinica Institute of Astronomy and Astrophysics and is funded by the Smithsonian Institution and the Academia Sinica.
SPL  acknowledges grants from the Ministry of Science and Technology of Taiwan 109-2112-M-007-010-MY3.
ZYL is supported in part by NASA 80NSSC18K1095 and NSF AST-1716259 and 1815784.

%

\vspace{5mm}
\facilities{ALMA, SMA}


\software{APLpy \citep{ro.br12}, astropy \citep{astr13, astr18}, CASA \citep{mcmu07}, MIR (https://github.com/qi-molecules/sma-mir), MIRIAD \citep{saul95}, ptemcee \citep{fore13, vous16}}



\clearpage

\appendix

\section{Entire moment maps of the CO isotopologue lines} \label{sec:mom01}
\begin{figure}[htbp]
\gridline{
\fig{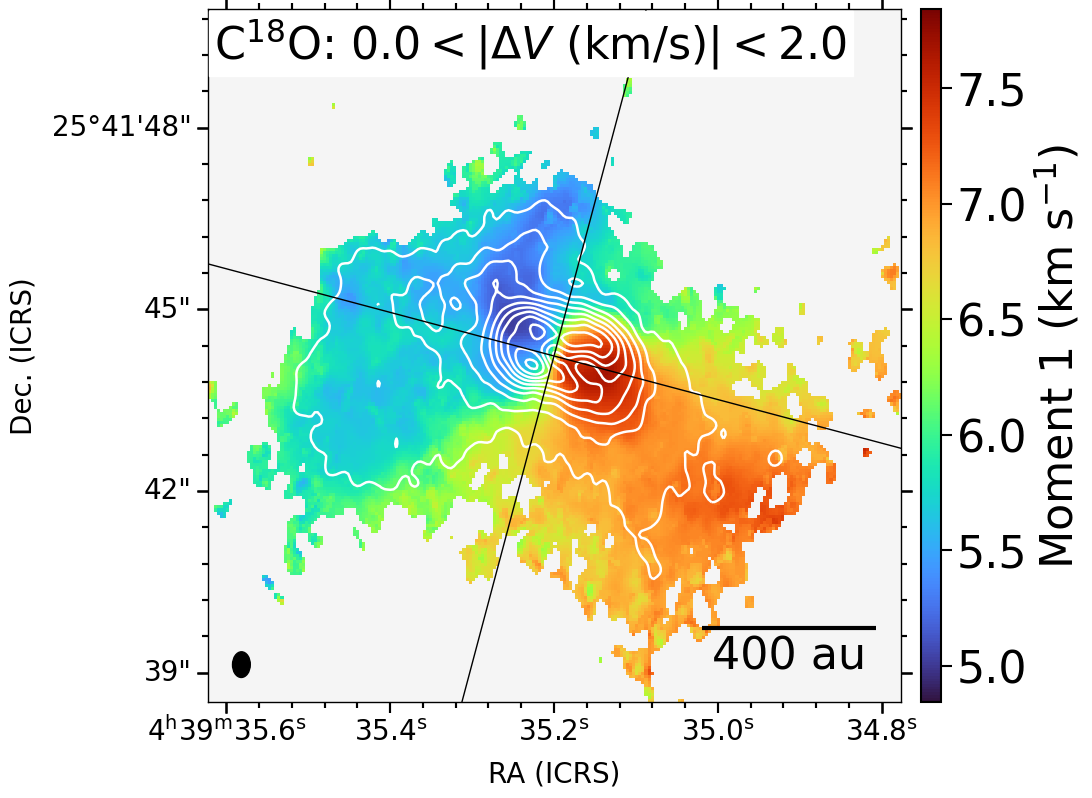}{0.3\textwidth}{(a)}
\fig{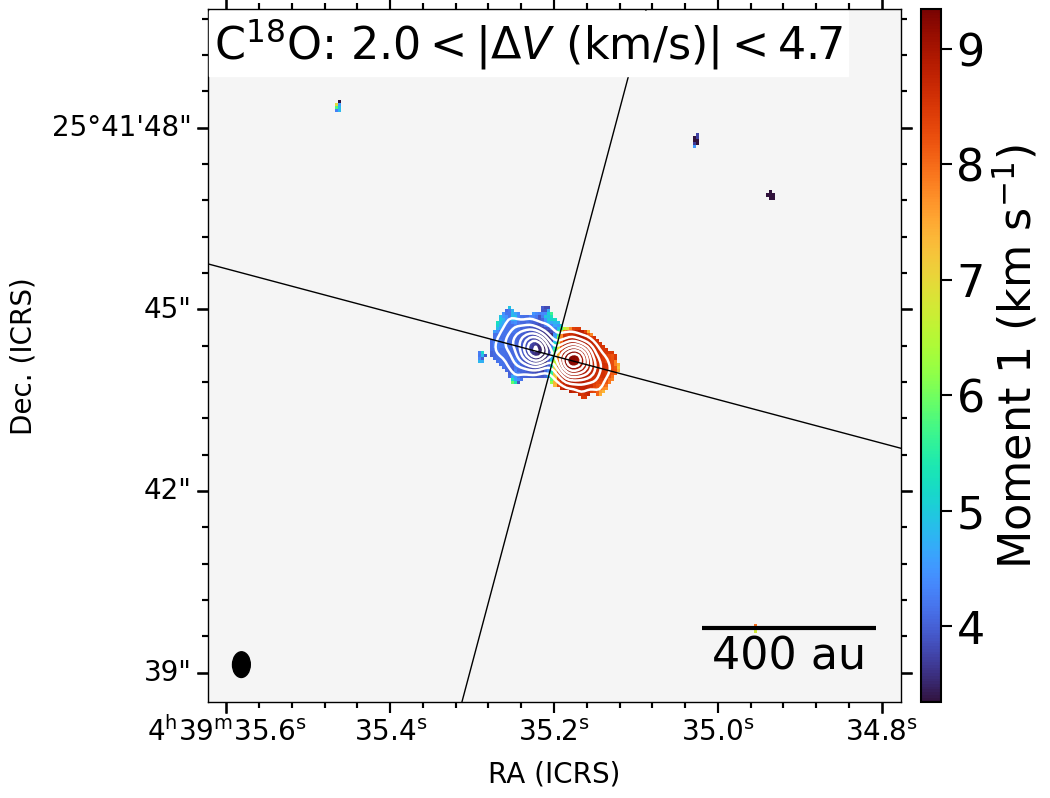}{0.3\textwidth}{(b)}
}
\gridline{
\fig{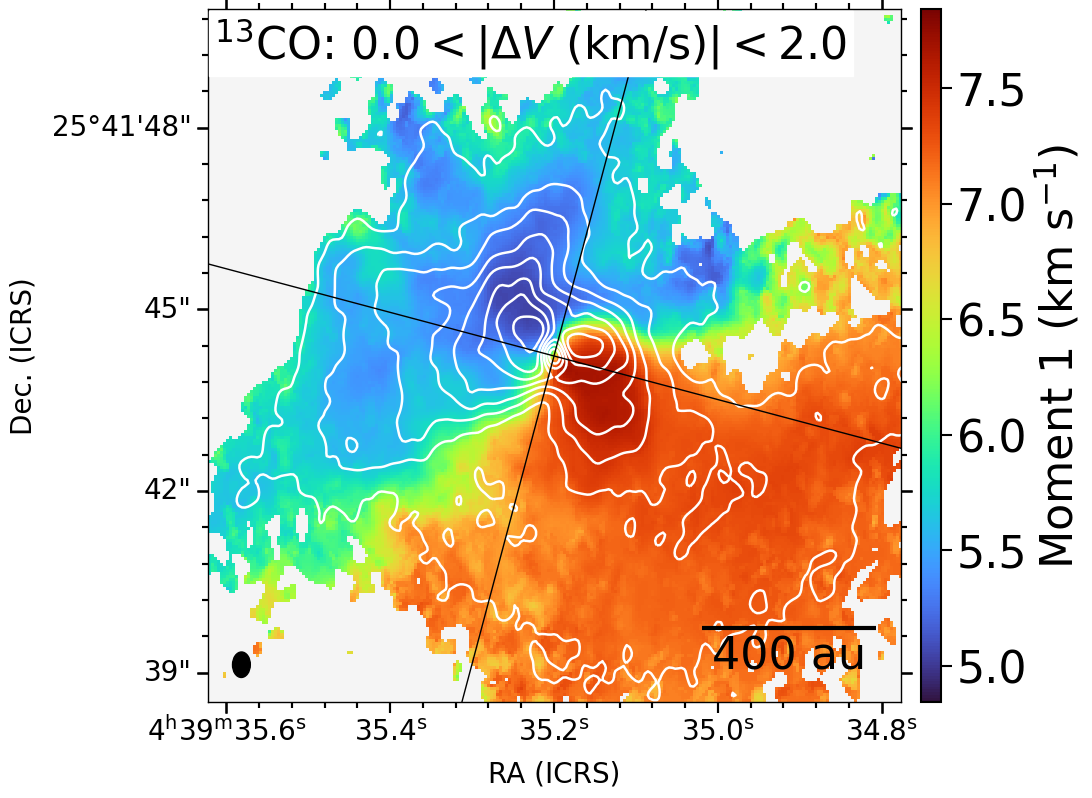}{0.3\textwidth}{(c)}
\fig{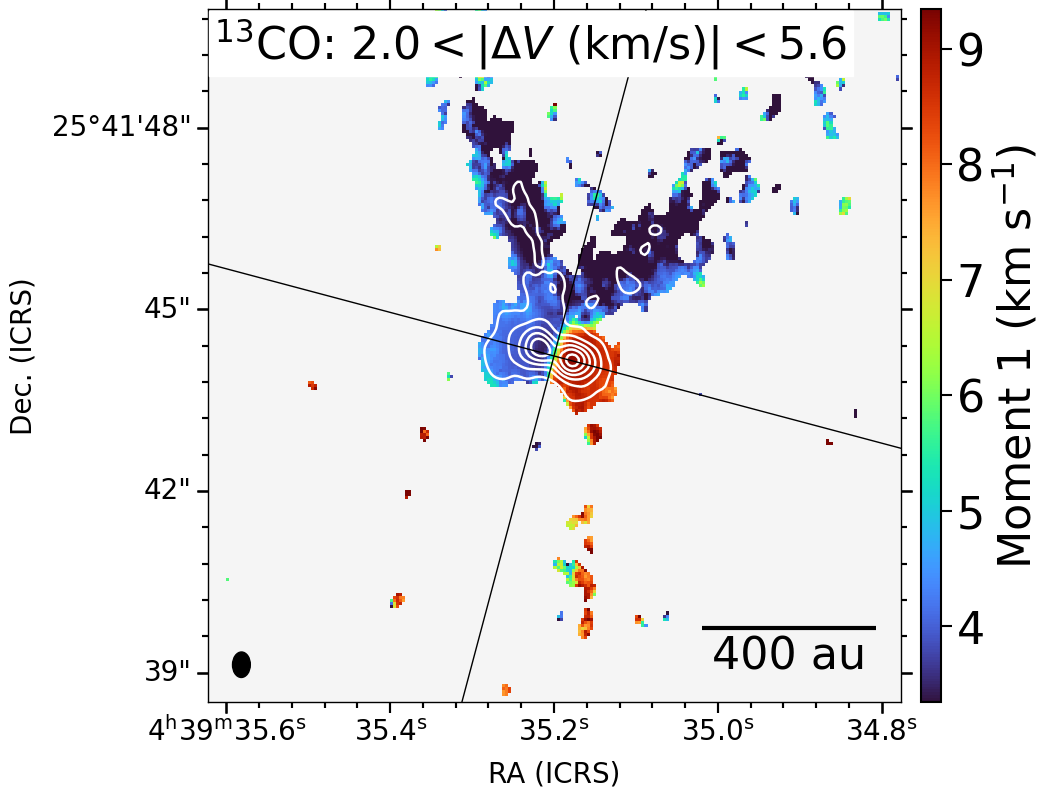}{0.3\textwidth}{(d)}
}
\gridline{
\fig{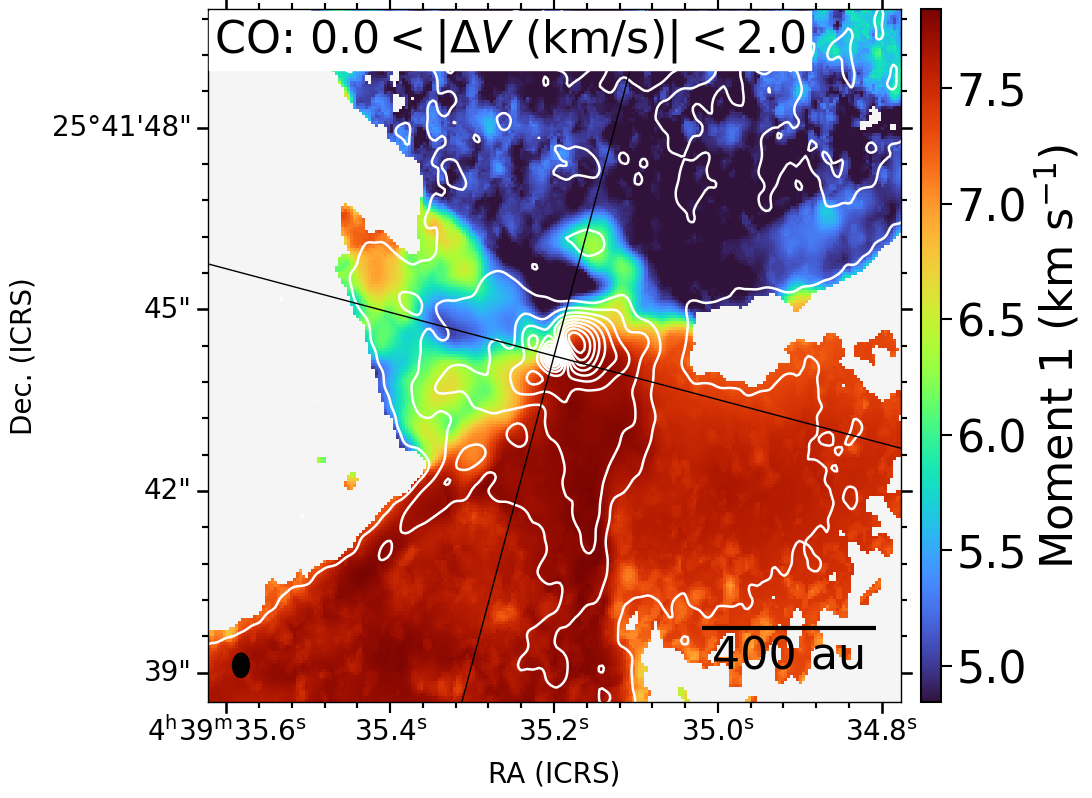}{0.3\textwidth}{(e)}
\fig{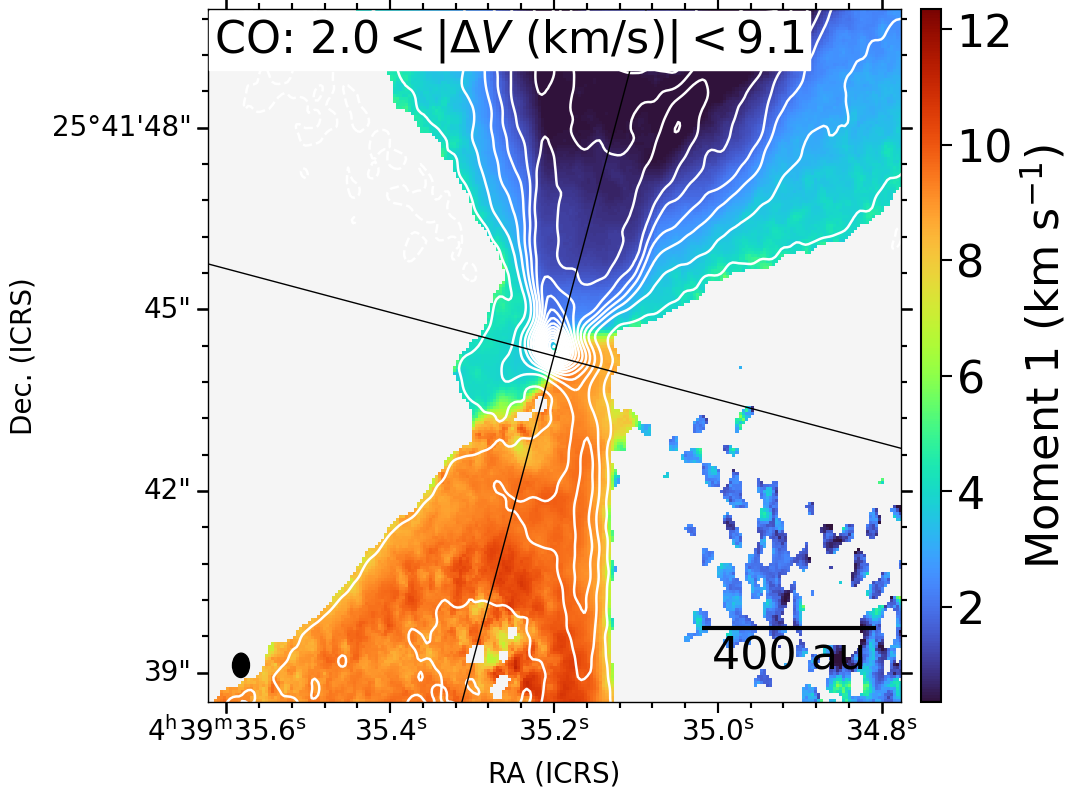}{0.3\textwidth}{(f)}
\fig{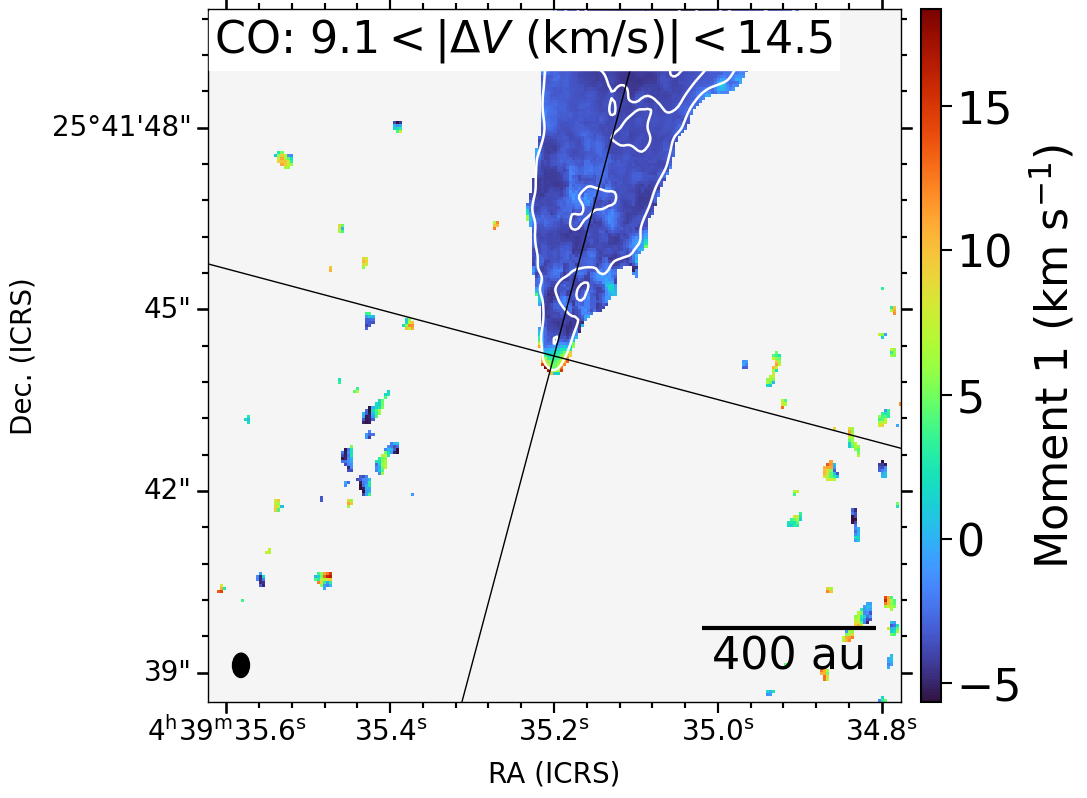}{0.3\textwidth}{(g)}
}
\caption{Integrated intensity (moment 0) and mean velocity (moment 1) maps of the CO isotopologue lines observed with ALMA. The spatial scale is four times larger than in Figure \ref{fig:line}. The integrated velocity range relative to $V_{\rm sys}=6.34\ \kms$ (Section \ref{sec:kep}) is denoted in each panel. Contour levels are in $12\sigma$, $24\sigma$, and $36\sigma$ steps for C$^{18}$O, $^{13}$CO and CO, respectively, from $12\sigma$, where $1\sigma$ corresponds to (a) 1.3, (b) 1.4, (c) 1.3, (d) 1.7, (e) 1.3, (f) 2.4, and (g) 2.1 $\mJB~\kms$. The moment 1 maps are masked at the $6\sigma$ levels. The diagonal lines denote the major (P.A.$=75\arcdeg$) and minor ($165\arcdeg$) axes of the TMC-1A disk (Section \ref{sec:sym}), centered at the protostellar position. The filled ellipse at the bottom-left corner of each panel denotes the ALMA synthesized beam.
\label{fig:mom01}}
\end{figure}

Figure \ref{fig:mom01} shows the CO isotopologue line emission in the $J=2-1$ transition on the entire envelope scale. The contour maps show the integrated intensity (moment 0), while the color maps show the mean velocity (moment 1). The integrated velocity range is the same as in Figure \ref{fig:line}. The spatial scale of the C$^{18}$O and $^{13}$CO maps is four times larger than in Figure \ref{fig:line}. These figures have angular resolutions three times higher than our previous observations toward TMC-1A reported in \citep{aso15}. Overall velocity structures of the C$^{18}$O envelope and those of the CO outflow are consistent with those in the lower-resolution observation in the same lines. \citet{aso15} did not observe the $^{13}$CO line, which traces the envelope and a part of the outflow. The higher angular resolution has revealed an excess in both moment 0 and moment 1 in a northern part in the C$^{18}$O and $^{13}$CO maps (Figure \ref{fig:mom01}a and \ref{fig:mom01}c). The CO emission, particularly in Figure \ref{fig:mom01}f, is stronger on the eastern, blueshifted side and on the western, redshifted side than the other side. This is consistent with another previous observation at an 8-au resolution \citep{bjer16}. 

\section{Comparison of the Keplerian Disk Model} \label{sec:chmod}
\begin{figure}[htbp]
\epsscale{1.115}
\plotone{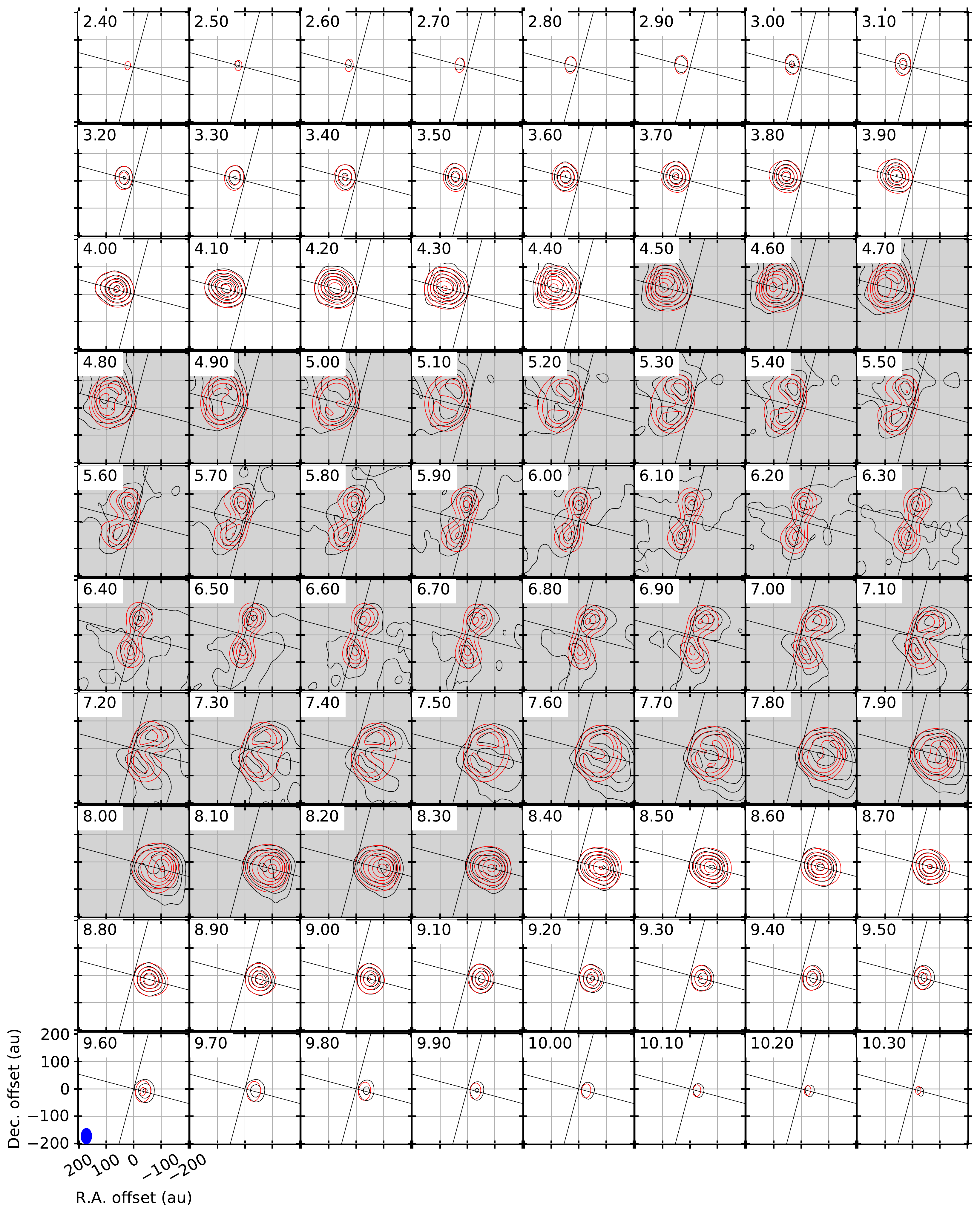}
\caption{The best-fit Keplerian disk model in Section \ref{sec:kep} (red) with the observed C$^{18}$O channel maps (black). The contour levels are in $10\sigma$ steps, where $1\sigma$ corresponds to $2~\mJB$. The filled ellipse is the C$^{18}$O synthesized beam. The velocity of each channel is denoted on the top-left corner of each panel. The diagonal lines are the same as those in Figure \ref{fig:almacont}. The channels with gray background are excluded from the fitting for the Keplerian disk in Section \ref{sec:kep}.
\label{fig:obsmodchan}}
\end{figure}

Figure \ref{fig:obsmodchan} shows comparison between the best-fit Kepleriam disk model derived in Section \ref{sec:kep} and the observed C$^{18}$O channel maps. The best-fit model reproduces the overall distribution of the C$^{18}$O emission observed with ALMA in the range $|V-V_{\rm sys}|>2.0~\kms$ (the channels with white background in Figure \ref{fig:obsmodchan}).

\begin{thebibliography}{}

\bibitem[Andrews \& Williams(2005)]{andr05} Andrews, S.~M. \& Williams, J.~P.\ 2005, \apj, 631, 1134. doi:10.1086/432712
\bibitem[Aso et al.(2015)]{aso15} Aso, Y., Ohashi, N., Saigo, K., et al.\ 2015, \apj, 812, 27. doi:10.1088/0004-637X/812/1/27
\bibitem[Astropy Collaboration et al.(2018)]{astr18} Astropy Collaboration, Price-Whelan, A.~M., Sip{\H{o}}cz, B.~M., et al.\ 2018, \aj, 156, 123. doi:10.3847/1538-3881/aabc4f
\bibitem[Astropy Collaboration et al.(2013)]{astr13} Astropy Collaboration, Robitaille, T.~P., Tollerud, E.~J., et al.\ 2013, \aap, 558, A33. doi:10.1051/0004-6361/201322068
\bibitem[Bjerkeli et al.(2016)]{bjer16} Bjerkeli, P., van der Wiel, M.~H.~D., Harsono, D., et al.\ 2016, \nat, 540, 406. doi:10.1038/nature20600
\bibitem[Chandler et al.(1998)]{chan98} Chandler, C.~J., Barsony, M., \& Moore, T.~J.~T.\ 1998, \mnras, 299, 789. doi:10.1046/j.1365-8711.1998.01818.x
\bibitem[Chandrasekhar \& Fermi(1953)]{ch.fe53} Chandrasekhar, S. \& Fermi, E.\ 1953, \apj, 118, 113. doi:10.1086/145731
\bibitem[Cho \& Lazarian(2007)]{ch.la07} Cho, J. \& Lazarian, A.\ 2007, \apj, 669, 1085. doi:10.1086/521805
\bibitem[Dapp et al.(2012)]{dapp12} Dapp, W.~B., Basu, S., \& Kunz, M.~W.\ 2012, \aap, 541, A35. doi:10.1051/0004-6361/201117876
\bibitem[Davis(1951)]{davi51} Davis, L.\ 1951, Physical Review, 81, 890. doi:10.1103/PhysRev.81.890.2
\bibitem[Foreman-Mackey et al.(2013)]{fore13} Foreman-Mackey, D., Hogg, D.~W., Lang, D., et al.\ 2013, \pasp, 125, 306. doi:10.1086/670067
\bibitem[Galli et al.(2018)]{gall18} Galli, P.~A.~B., Loinard, L., Ortiz-L{\'e}on, G.~N., et al.\ 2018, \apj, 859, 33. doi:10.3847/1538-4357/aabf91
\bibitem[Girart et al.(2018)]{gira18} Girart, J.~M., Fern{\'a}ndez-L{\'o}pez, M., Li, Z.-Y., et al.\ 2018, \apjl, 856, L27. doi:10.3847/2041-8213/aab76b
\bibitem[Gold(1952)]{gold52} Gold, T.\ 1952, \mnras, 112, 215. doi:10.1093/mnras/112.2.215
\bibitem[Harsono et al.(2018)]{hars18} Harsono, D., Bjerkeli, P., van der Wiel, M.~H.~D., et al.\ 2018, Nature Astronomy, 2, 646. doi:10.1038/s41550-018-0497-x
\bibitem[Harsono et al.(2014)]{hars14} Harsono, D., J{\o}rgensen, J.~K., van Dishoeck, E.~F., et al.\ 2014, \aap, 562, A77. doi:10.1051/0004-6361/201322646
\bibitem[Heitsch et al.(2001)]{heit01} Heitsch, F., Zweibel, E.~G., Mac Low, M.-M., et al.\ 2001, \apj, 561, 800. doi:10.1086/323489
\bibitem[Hennebelle \& Ciardi(2009)]{he.ci09} Hennebelle, P. \& Ciardi, A.\ 2009, \aap, 506, L29. doi:10.1051/0004-6361/200913008
\bibitem[Hoang et al.(2018)]{hoan18} Hoang, T., Cho, J., \& Lazarian, A.\ 2018, \apj, 852, 129. doi:10.3847/1538-4357/aa9edc
\bibitem[Hull et al.(2017a)]{hull17a} Hull, C.~L.~H., Girart, J.~M., Tychoniec, {\L}., et al.\ 2017, \apj, 847, 92. doi:10.3847/1538-4357/aa7fe9
\bibitem[Hull et al.(2017b)]{hull17b} Hull, C.~L.~H., Mocz, P., Burkhart, B., et al.\ 2017, \apjl, 842, L9. doi:10.3847/2041-8213/aa71b7
\bibitem[Inutsuka et al.(2010)]{inut10} Inutsuka, S.-. ichiro ., Machida, M.~N., \& Matsumoto, T.\ 2010, \apjl, 718, L58. doi:10.1088/2041-8205/718/2/L58
\bibitem[Joos et al.(2012)]{joos12} Joos, M., Hennebelle, P., \& Ciardi, A.\ 2012, \aap, 543, A128. doi:10.1051/0004-6361/201118730
\bibitem[Kataoka et al.(2012)]{kata12} Kataoka, A., Machida, M.~N., \& Tomisaka, K.\ 2012, \apj, 761, 40. doi:10.1088/0004-637X/761/1/40
\bibitem[Kataoka et al.(2016)]{kata16} Kataoka, A., Muto, T., Momose, M., et al.\ 2016, \apj, 820, 54. doi:10.3847/0004-637X/820/1/54
\bibitem[Kataoka et al.(2017)]{kata17} Kataoka, A., Tsukagoshi, T., Pohl, A., et al.\ 2017, \apjl, 844, L5. doi:10.3847/2041-8213/aa7e33
\bibitem[Krasnopolsky et al.(2011)]{kras11} Krasnopolsky, R., Li, Z.-Y., \& Shang, H.\ 2011, \apj, 733, 54. doi:10.1088/0004-637X/733/1/54
\bibitem[Kwon et al.(2019)]{kwon19} Kwon, W., Stephens, I.~W., Tobin, J.~J., et al.\ 2019, \apj, 879, 25. doi:10.3847/1538-4357/ab24c8
\bibitem[Lambrechts \& Johansen(2014)]{la.jo14} Lambrechts, M. \& Johansen, A.\ 2014, \aap, 572, A107. doi:10.1051/0004-6361/201424343
\bibitem[Lazarian(1994)]{laza94} Lazarian, A.\ 1994, \mnras, 268, 713. doi:10.1093/mnras/268.3.713
\bibitem[Lee et al.(2018)]{lee18} Lee, C.-F., Li, Z.-Y., Ching, T.-C., et al.\ 2018, \apj, 854, 56. doi:10.3847/1538-4357/aaa769
\bibitem[Le Gouellec et al.(2019)]{lego19} Le Gouellec, V.~J.~M., Hull, C.~L.~H., Maury, A.~J., et al.\ 2019, \apj, 885, 106. doi:10.3847/1538-4357/ab43c2
\bibitem[McMullin et al.(2007)]{mcmu07} McMullin, J.~P., Waters, B., Schiebel, D., et al.\ 2007, Astronomical Data Analysis Software and Systems XVI, 376, 127
\bibitem[Mellon \& Li(2008)]{me.li08} Mellon, R.~R. \& Li, Z.\ 2008, American Astronomical Society Meeting Abstracts \#211
\bibitem[Ohashi \& Kataoka(2019)]{ohas19} Ohashi, S. \& Kataoka, A.\ 2019, \apj, 886, 103. doi:10.3847/1538-4357/ab5107
\bibitem[Ohashi et al.(2020)]{ohas20} Ohashi, S., Kobayashi, H., Nakatani, R., et al.\ 2020, arXiv:2012.04082
\bibitem[Ostriker et al.(2001)]{ostr01} Ostriker, E.~C., Stone, J.~M., \& Gammie, C.~F.\ 2001, \apj, 546, 980. doi:10.1086/318290
\bibitem[Padoan et al.(2001)]{pado01} Padoan, P., Goodman, A., Draine, B.~T., et al.\ 2001, \apj, 559, 1005. doi:10.1086/322504
\bibitem[Robitaille \& Bressert(2012)]{ro.br12} Robitaille, T. \& Bressert, E.\ 2012, Astrophysics Source Code Library. ascl:1208.017
\bibitem[Sault et al.(1995)]{saul95} Sault, R.~J., Teuben, P.~J., \& Wright, M.~C.~H.\ 1995, Astronomical Data Analysis Software and Systems IV, 77, 433
\bibitem[Takakuwa et al.(2014)]{taka14} Takakuwa, S., Saito, M., Saigo, K., et al.\ 2014, \apj, 796, 1. doi:10.1088/0004-637X/796/1/1
\bibitem[Tobin et al.(2016)]{tobi16} Tobin, J.~J., Kratter, K.~M., Persson, M.~V., et al.\ 2016, \nat, 538, 483. doi:10.1038/nature20094
\bibitem[Tomida et al.(2015)]{tomi15} Tomida, K., Okuzumi, S., \& Machida, M.~N.\ 2015, \apj, 801, 117. doi:10.1088/0004-637X/801/2/117
\bibitem[Tsukamoto et al.(2015)]{tsuk15} Tsukamoto, Y., Iwasaki, K., Okuzumi, S., et al.\ 2015, \apjl, 810, L26. doi:10.1088/2041-8205/810/2/L26
\bibitem[Vousden et al.(2016)]{vous16} Vousden, W.~D., Farr, W.~M., \& Mandel, I.\ 2016, \mnras, 455, 1919. doi:10.1093/mnras/stv2422
\bibitem[Wilner \& Welch(1994)]{wi.we94} Wilner, D.~J., \& Welch, W.~J.\ 1994, \apj, 427, 898 
\bibitem[Yang et al.(2017)]{yang17} Yang, H., Li, Z.-Y., Looney, L.~W., et al.\ 2017, \mnras, 472, 373. doi:10.1093/mnras/stx1951
\bibitem[Yang et al.(2016)]{yang16} Yang, H., Li, Z.-Y., Looney, L.~W., et al.\ 2016, \mnras, 460, 4109. doi:10.1093/mnras/stw1253
\bibitem[Zhao et al.(2018)]{zhao18} Zhao, B., Caselli, P., Li, Z.-Y., et al.\ 2018, \mnras, 473, 4868. doi:10.1093/mnras/stx2617


\end{thebibliography}



\end{document}